\documentclass[10pt,prb,aps,superscriptaddress,amsmath,amssymb,twocolumn,floatfix,longbibliography]{revtex4-1}
\usepackage[pdftex]{graphicx}
\usepackage{bm}
\usepackage{bbold}
\usepackage{mathtools}
\usepackage{nccmath}
\usepackage[dvipsnames]{xcolor}
\usepackage[bookmarksopen,colorlinks,linkcolor=blue,citecolor=olive,urlcolor=red]{hyperref}

\newcommand{\bg}{\begin{gather}}
\newcommand{\eg}{\end{gather}}

\newcommand{\be}{\begin{equation}}
\newcommand{\ee}{\end{equation}}

\newcommand{\corr}[1]{\langle #1\rangle}

\newcommand{\str}{\mathop{\rm str}}
\newcommand{\sdet}{\operatorname{sdet}}

\newcommand{\diag}{\operatorname{diag}}

\newcommand{\var}{\operatorname{var}}
\newcommand{\cum}[1]{\langle \langle #1^3 \rangle \rangle}

\newcommand{\thone}{\theta_\text{B1}}
\newcommand{\thtwo}{\theta_\text{B2}}
\newcommand{\thF}{\theta_\text{F}}

\newcommand{\aone}{a_\text{B1}}
\newcommand{\atwo}{a_\text{B2}}
\newcommand{\aF}{a_\text{F}}

\newcommand{\gavg}{\langle g \rangle}

\begin{document}

\title{Anderson localization at the boundary of a two-dimensional topological superconductor}

\author{Daniil S.\ Antonenko}
\affiliation{Department of Physics, Drexel University, Philadelphia, PA 19104, USA}
\affiliation{L.\ D.\ Landau Institute for Theoretical Physics, Chernogolovka 142432, Russia}

\author{Eslam Khalaf}
\affiliation{Department of Physics, Harvard University, Cambridge, MA 02138, USA}

\author{Pavel M.\ Ostrovsky}
\affiliation{L.\ D.\ Landau Institute for Theoretical Physics, Chernogolovka 142432, Russia}
\affiliation{Max-Planck-Institut f{\"u}r Festk{\"o}rperforschung, 70569 Stuttgart, Germany}

\author{Mikhail A.\ Skvortsov}
\affiliation{L.\ D.\ Landau Institute for Theoretical Physics, Chernogolovka 142432, Russia}

\date{\today}

\begin{abstract}
A one-dimensional boundary of a two-dimensional topological superconductor can host a number of topologically protected chiral modes. Combining two topological superconductors with different topological indices, it is possible to achieve a situation when only a given number of channels ($m$) are topologically protected, while others are not and therefore are subject to Anderson localization in the presence of disorder. We study transport properties of such quasi-one-dimensional quantum wires with broken time-reversal and spin-rotational symmetries (class D) and calculate the average conductance, its variance and the third cumulant, as well as the average shot noise power. The results are obtained for arbitrary wire length, tracing a crossover from the diffusive Drude regime to the regime of strong localization where only $m$ protected channels conduct. Our approach is based on the non-perturbative treatment of the non-linear supersymmetric sigma model of symmetry class D with two replicas developed in our recent publication [D. S. Antonenko \emph{et al.}, Phys. Rev. B \textbf{102}, 195152 (2020)]. The presence of topologically protected modes results in the appearance of a topological Wess-Zumino-Witten term in the sigma-model action, which leads to an additional subsidiary series of eigenstates of the transfer-matrix Hamiltonian. The developed formalism can be applied to study the interplay of Anderson localization and topological protection in quantum wires of other symmetry classes.
\end{abstract}

\maketitle

\section{Introduction}

Quantum localization is a fundamental phenomenon first introduced in the seminal paper by Anderson \cite{Anderson_first} over 60 years ago. Since then numerous types and forms of localization effects have been identified \cite{EversMirlin} ranging from the quantum Hall effect \cite{GirvinPrange} along with other topological insulators and superconductors \cite{HasanKane} to many-body localization physics of interacting systems \cite{MBL}. It is now a common knowledge that disorder can dramatically change the nature of quasiparticle wavefunctions leading to exponential suppression of transport due to quantum interference. These effects are especially pronounced in low-dimensional systems, where quasiparticles have less volume to avoid impurity scattering. In particular, it has been argued by a general scaling analysis that localization is unavoidable in one-dimensional systems \cite{band4}. However, this statement in its general form is undermined by recent developments in the physics of topological systems \cite{HasanKane}. It was demonstrated that one-dimensional conductors of certain symmetry can host topologically protected modes that evade localization as long as the symmetry is preserved in the presence of disorder \cite{Brouwer_delocalisation, BagretsKamenev, Akhmerov_classD}. Such topologically protected modes can coexist with usual unprotected modes, when only the latter are subject to localization \cite{KSO2016}. In the present paper, we consider a particular model of a superconductor with both protected and unprotected conducting channels and study various transport characteristics in this mixed arrangement for a broad range of system lengths.

\begin{figure}[b]
\vskip-2mm
\includegraphics[width=0.8\linewidth]{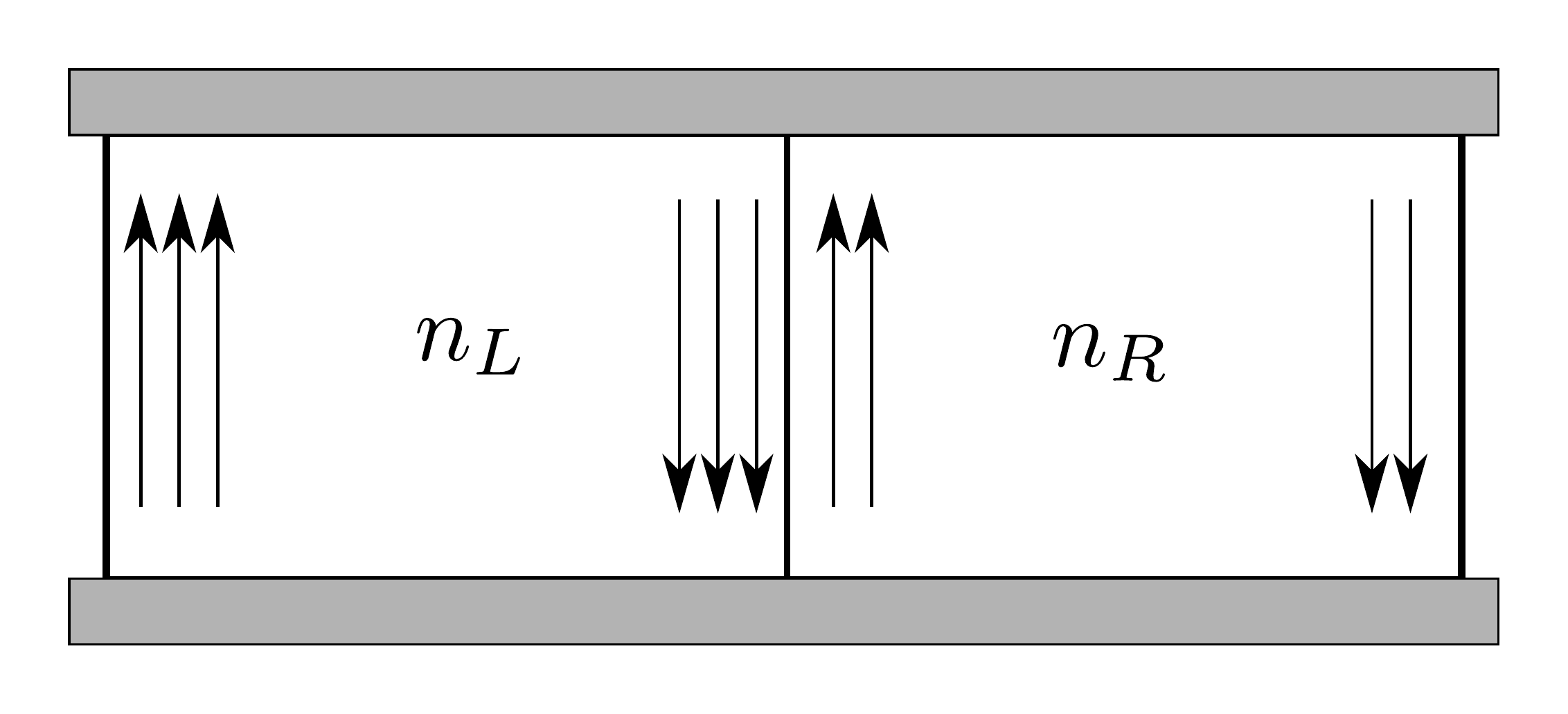}
\vskip-2mm
\caption{Junction of two 2D topological superconductors as a possible experimental implementation of a quasi-1D system possessing $\left| n_R - n_L \right|$ protected modes and $\min\{n_L, n_R\}$ unprotected modes at the interface. Here $n_L$ and $n_R$ denote topological indexes of left and right superconductors, correspondingly. Terminals for transport measurements are shown by gray rectangles.}
\label{fig:edge}
\end{figure}

We study transport properties of the one-dimenstional (1D) boundary states of a two-dimensional (2D) topological superconductor of symmetry class D (broken time-reversal and spin-rotational symmetries). According to the general classification of topological insulators and superconductors \cite{Kitaev2009, 10foldway}, in this symmetry class 2D bulk is characterized by a $\mathbb{Z}$ topological index. Due to the bulk-boundary correspondence, the same index provides the number of topologically protected chiral edge states circulating around the boundary of the system. We consider a general setup with both protected and unprotected channels, which can be realized in a junction of two such 2D topological superconductors with different values of their topological indexes as shown in Fig.\ \ref{fig:edge}. This difference induces an imbalance between right- and left-propagating channels at the interface between superconductors thus leading to the coexistence of protected and unprotected modes \cite{KSO2016}.

Probing quasiparticle edge transport is a challenging task in a topological superconductor since electrical transport measurements are compromised by the shunting effect of the superconducting bulk. Spin of edge excitations is also not conserved due to the lack of spin-rotational symmetry making their spin conductance an ill-defined quantity. The only conserved physical property of the edge modes is their energy, hence it is only the \emph{thermal conductance} that can reliably quantify edge transport. This is why topological effects in the superconductors of class D were dubbed as thermal quantum Hall effect (TQHE) \cite{TQHE}. We will characterize the edge modes by the value of their thermal conductance $G_T$ and introduce a corresponding dimensionless conductance $g = G_T / G_0$, where $G_0 = \pi k_B^2 T / 6 \hbar$ is the quantum of thermal conductance \cite{ReadGreen_2000}. To simplify the analysis, we will limit our consideration to the case of low temperatures and hence neglect possible inelastic scattering.

One particular way to controllably create a quasi-1D system with both protected and unprotected modes is bringing into contact two 2D topological superconductors with different topological indexes $n_L$ and $n_R$ as shown in Fig.\ \ref{fig:edge}. In this way, we obtain unequal numbers of left- and right-propagating modes at the interface. The difference $n_R - n_L$ gives the number of topologically protected modes that evade localization \cite{KSO2016}. A similar setup was implemented experimentally for the quantum Hall physics (i.e.\ for a unitary symmetry class) in a 2D electron gas, located at a surface of a crystal with different filling factors on adjacent faces \cite{Grayson2005_edges, Grayson2007_edges, Grayson2008_edges, Steinke_edges}. In our case of a topological superconductor, the topological classification and the geometry of protected edge states is the same.

In the setup of Fig.\ \ref{fig:edge}, the total conductance does not depend on the direction of the current between the terminals (denoted by gray rectangles). However, the current is distributed differently between the inner interface and outer edges depending on the overall current direction. Assume for a moment $n_L > n_R$. Then at the interface there are $m = n_L - n_R$ topologically protected modes propagating downwards. When a current flows from the upper to the lower terminal, it is carried by these modes together with $n_R$ modes at the rightmost boundary. In total there are $n_L$ protected modes and $n_R$ unprotected channels carrying this current. In the opposite case when the current flows upwards it is carried by $n_L$ protected modes on the leftmost edge of the system together with $n_R$ unprotected channels at the interface in the middle. In the opposite case $n_L < n_R$ the argument is similar. Hence there are always $\max\{n_L, n_R\}$ protected and $\min\{n_L, n_R\}$ unprotected channels irrespective of the current direction. All unprotected modes are at the interface between superconductors while the number of protected modes at the same interface is $m = \left| n_L - n_R \right|$.

To simplify the analysis we find it convenient to consider \emph{direction-averaged} dimensionless conductance of the interface $g$. We decompose the overall conductance $g_\text{tot}$ into the contributions of outer edges and inner interface with both quantities averaged over the two possible directions of the current \cite{KSO2016}:
\be
\label{g_tot}
	g_\text{tot} = \frac{n_L + n_R}{2} + g.
\ee
Here the first term corresponds to the direction-averaged contribution of the outer edges of the junction (see Fig.~\ref{fig:edge}). In the following we focus only on the second term $g$. We will study statistical properties of the interface conductance $g$, in particular, its averaged value $\gavg$, variance $\var g$ and the third cumulant $\cum g$. Also, we will compute the shot noise power of quasiparticles which is also due to the channels at the interface.

The case without protected modes ($m=0$, corresponding to $n_L = n_R$ in the aforementioned setup) is special and was extensively studied previously \cite{Beenaker_Dorokhov, BagretsKamenev, classD_first}. It can be implemented in a superconducting wire with broken time-reversal and spin-rotational symmetries and does not require an adjacent two-dimensional gapped bulk superconductor. Such systems have attracted a great interest in recent years both theoretically and experimentally since they are very promising candidates to realize Majorana bound states \cite{Kitaev_chain, Oreg, Beenakker-Majorana}. Gapped phases of such disordered wires are characterised by a one-dimensional $\mathbb{Z}_2$ topological index $q$, which takes only two possible values $0$ or $1$. This index distinguishes a trivial gapped state $q = 0$ and a topological state $q = 1$ that carries a pair of zero-energy Majorana states at the ends of the wire. Critical state between these two gapped phases is not fully localized by disorder \cite{Akhmerov_classD, Brouwer_delocalisation, GruzbergReadVishveshwara, BagretsKamenev}. Instead it is characterized by subohmic scaling of the average conductance with the wire length $\gavg \propto 1/ \sqrt{L}$.

In our previous publication \cite{classD_first}, we computed conductance variance and other higher transport moments for a disordered Majorana wire with many conducting channels fine-tuned to the critical regime. In the present paper we extend our research taking into account possible topologically protected modes, which can appear only if the studied 1D system is realized at an edge of a two-dimensional topological superconductor. Since such an edge model is gapless, the $\mathbb{Z}_2$ topological classification of Majorana wires does not apply in this case.

We use the formalism of the supersymmetric nonlinear sigma model \cite{Efetov-book}, which is applicable to disordered systems with the diffusive regime at moderately short lengths. In a 1D geometry this implies a large number of conducting modes $n_{L,R} \gg 1$.

While the symmetry class of the sigma model is determined by the symmetry of the underlying system, the rank of the model depends on the particular averaged quantity to be studied. The simplest quantities such as the average density of states or conductance involve at most two Green functions (retarded and advanced). They can be calculated within the minimal sigma model of rank $n = 1$. Such analysis for a disordered 1D system of the symmetry class D was carried out both in the topologically trivial ($m = 0$) \cite{Zirnbauer1991, Zirnbauer1992, BocquetZirnbauer, BagretsKamenev} and topologically nontrivial ($m \neq 0$) \cite{KSO2016, Eslam_thesis} settings. In the latter case, inclusion of the channel imbalance results in the appearance of a topological Wess-Zumino-Witten (WZW) term in the sigma-model action, with the coefficient proportional to $m = |n_L - n_R|$.

The minimal one-replica sigma model of class D allows to calculate only the average conductance $\gavg$. Higher moments of transport, such as conductance variance or shot noise power require averaging of more than two Green functions. Hence the sigma model with at least $n = 2$ replicas is required. Such a model was developed in our earlier work \cite{classD_first} for the topologically trivial wires with $m = 0$. In the present paper we generalize this model to include possible topologically protected modes in the case $m \neq 0$. Using the generalized model we will calculate the first three moments of the conductance as well as the average Fano factor for the thermal shot noise.

\begin{figure}
\includegraphics[width=\linewidth]{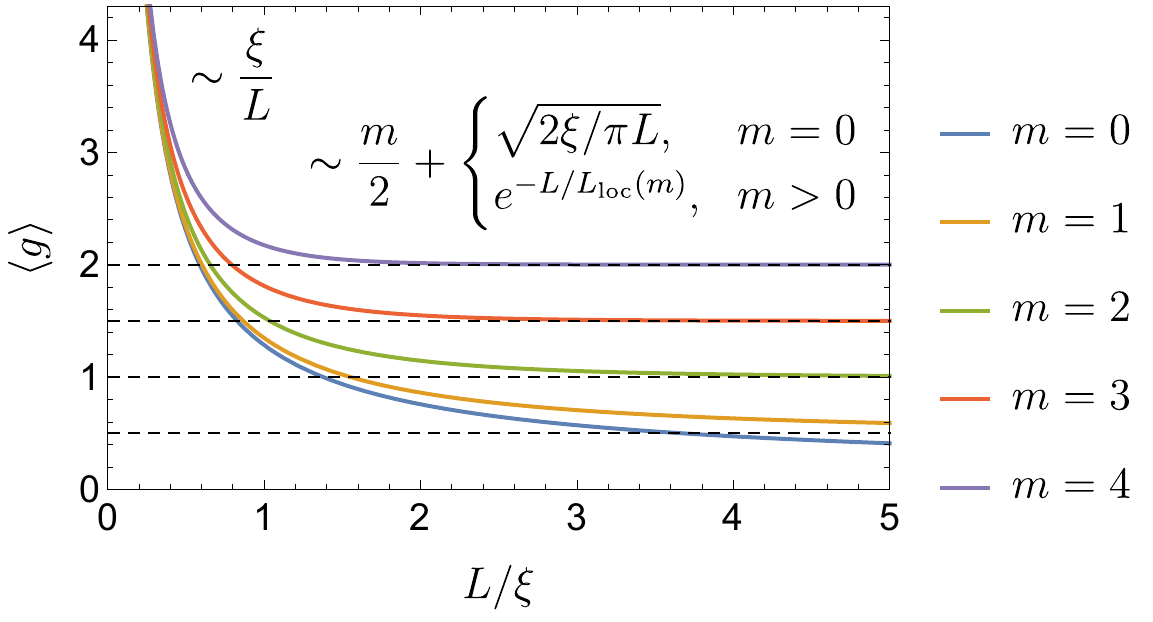}
\caption{Average conductance $\gavg$ in the presence of $m$ protected topological modes as a function of the wire length $L$. Dashed lines show the leading short- and long-wire asymptotics. Localization length of unprotected modes $L_\text{loc}(m)$ decreases with the increase of $m$, following Eq.~\eqref{loc-length}.}
\label{fig:conductance}
\end{figure}

Anticipating detailed discussion of our results in Secs.\ \ref{sec:results} and \ref{sec:conclusion}, in Fig.\ \ref{fig:conductance} we present the average conductance of a 1D system with $m$ topologically protected modes as a function of the wire length $L$. It shows how the classical ohmic behaviour $\gavg \sim \xi / L$ at short lengths ($\xi$ is an analogue of the localization length) undergoes a crossover to the regime where all unprotected modes are fully localized. In the limit $L \gg \xi$ conductance approaches the value $\gavg = m/2$ since only $m$ topologically protected modes evade Anderson localization  and they carry current only in one direction. For $m>0$, correction to this asymptotic value is exponential, $\corr{g}-m/2\propto e^{-L/L_\text{loc}(m)}$, with the localization length of unprotected modes $L_\text{loc}(m)$ given by
\be
\label{loc-length}
  L_\text{loc}(m)/\xi =
  \begin{cases}
    8/m^2 , & m\leq 4 ; \\
    1/(m-2) , & m\geq 4 .
  \end{cases}
\ee
It is worth noting that $L_\text{loc}(m)$ decreases with increasing $m$, as can be seen in Fig.\ \ref{fig:conductance}.

The main technical achievement of this paper is the construction of the full set of eigenfunctions for the transfer-matrix Hamiltonian of the two-replica supersymmetric sigma model with the WZW term. This technique is closely connected with the Fourier analysis on the corresponding symmetric superspace (sigma-model manifold). In our previous work, we have constructed a full set of eigenstates of the Laplace operator on this superspace \cite{classD_first}. This set has a complex hierarchical structure with some parts inherited from the one-replica model. In the present paper, we generalize this construction to the $m \neq 0$ case, which results in a vector potential in the transfer-matrix Hamiltonian corresponding to the WZW term in the action.
We will demonstrate that this vector potential gives rise to additional eigenfunctions corresponding to bound states on the supermanifold.

The paper is organized as follows. In Sec.\ \ref{sec:technical} we introduce the sigma model formalism in application to the two-replica version of class D and present the calculation scheme for the transport moments. We briefly outline the algorithm described in detail in Ref.\ \onlinecite{classD_first} and explain the modifications introduced by the WZW term. In Sec.\ \ref{sec:eigenbasis} we construct the radial eigenbasis of the Laplace operator on the sigma-model supermanifold and discuss its differences compared to the $m = 0$ case. We use the constructed eigenfunctions to build the heat kernel of the sigma model in Sec.\ \ref{sec:heat_kernel} and comment on the integration peculiarities in the momentum space. We present results of our study in Sec.\ \ref{sec:results} and draw conclusions in Sec.~\ref{sec:conclusion}. An outline of the sigma-model derivation with the WZW term is given in Appendix \ref{appsigma}. Details of the gauge factor calculation for the vector potential induced by the WZW term can be found in Appendix \ref{app:gauge}.

\section{Sigma model formalism}
\label{sec:technical}

In this Section we introduce the nonlinear sigma-model formalism and apply it to the calculation of the transport properties. In many aspects our analysis follows the simpler $m = 0$ case (disordered wire without protected modes) and we refer the reader to Ref.\ \onlinecite{classD_first} for the details of such a calculation. In the present Section we will particularly focus on the points that are specific for the topologically nontrivial $m \neq 0$ case.

\subsection{Sigma-model with WZW term}

We perform our study in the framework of the supersymmetric nonlinear sigma model \cite{Efetov-book}. Its objective is to average a product of a certain number $n$ of bosonic and an equal number $n$ of fermionic Green functions over disorder realizations. We call the parameter $n$ the number of replicas in the theory. Any superconducting Hamiltonian has a special mirror symmetry of its spectrum that results in an additional structure of the sigma model in the particle-hole space.

By extracting low-energy properties via functional integration approach, we arrive at an effective description of soft electronic modes of the disordered system (diffusons and cooperons) in terms of a field theory for a supermatrix field $Q$, which belongs to a certain supermanifold. In our case, the supermatrix $Q \in \text{BF} \otimes \text{PH} \otimes \text{R}$ belongs to the tensor product of the Bose-Fermi (BF), particle-hole (PH), and replica (R) spaces. Hence the overall size of $Q$ is $4n$. In addition, $Q$ satisfies $Q^2 = 1$ and is subject to the charge conjugation constraint $\overline{Q} = C^T Q^T C = -Q$. The latter reflects superconducting symmetry of the underlying Bogoliubov-de Gennes Hamiltonian. The matrix $C$, which defines charge conjugation, should satisfy $C^T C = 1$ and $C^2 = - k$, where $k = \diag \{1, -1\}_\text{BF}$ is a signature matrix in the BF space [explicit representation of the matrix $C$ can be found in Appendix \ref{app:gauge}, see Eq.\ (\ref{C})]. In general, the matrix $Q$ can be represented as
\be
\label{Q_via_Lambda_and_T}
	Q = T^{-1} \Lambda T,
	\qquad
	\overline{T} =  T^{-1}.
\ee
The matrix $\Lambda$ represents a certain point on the sigma-model manifold satisfying $\overline\Lambda = - \Lambda$, $\str \Lambda = 0$, and $\Lambda^2 = \mathbb{1}$. The whole manifold is then generated by rotations of $\Lambda$ with elements $T$ from the corresponding supergroup $G$. Within $G$ there are certain rotations that commute with $\Lambda$ and hence do not change $Q$. They form a subgroup $K \subset G$. The sigma-model manifold is thus a \textit{symmetric superspace} (coset) $G / K$ with $[K, \Lambda] = 0$.

The sigma-model action for the quasi-1D topological wire of symmetry class D has the form \cite{KSO2016}
\be
\label{sigma_model_action}
	S = - \int_{0}^{L}dx\, \str \left[ \frac{\xi}{16} (\nabla Q)^{2} + \frac{m}{4} T^{-1} \Lambda \nabla T \right].
\ee
Here $\xi = (n_L + n_R) l$ is the correlation length and $l$ is the mean free path. The second, Wess-Zumino-Witten (WZW) term in the action takes into account the imbalance between right- and left-propagating channels. Its prefactor $m = \left| n_R - n_L \right|$ determines the number of topologically protected modes in the system. An outline of the derivation of Eq.\ (\ref{sigma_model_action}) is given in Appendix \ref{appsigma}.

For a supersymmetric sigma model of class D with $n$ replicas, the target manifold is $\text{SpO}(2n, \mathbb{R} | 2n) / \text{U}(n|n)$\cite{GruzbergMirlinZirnbauer_classification}. Its compact Fermi-Fermi sector $\text{O}(2n) / \text{U}(n)$ has two disjoint components corresponding to the two  components of the $\text{O}(2n)$ group. Such a structure of the sigma-model manifold implies that the $Q$ matrix configurations can include jumps between the two parts of the target manifold. Inclusion of such jumps is necessary to ensure proper exponential localization in wires of class D without topological terms \cite{BocquetZirnbauer, ReadLudwig2000, BagretsKamenev}.

In our earlier work, Ref.\ \onlinecite{classD_first}, we studied a critical regime between topological and trivial gapped states in a Majorana wire of class D. This problem corresponds to the case $m=0$ and to the absence of jumps between two parts of the manifold. The latter condition is ensured by the fact that the statistical weight of such jumps changes sign at the transition between trivial and Majorana state \cite{BagretsKamenev}. In the present case of an imbalanced system with $m \neq 0$, the jumps of the matrix $Q$ are also fully suppressed. Indeed, the WZW term introduces a random phase factor to the amplitude of every jump. Averaging over these phases completely cancels possible contribution of such discontinuous configurations of $Q$. This is quite natural since both in the critical regime with $m = 0$ studied earlier and for the topological edge transport with $m \neq 0$ considered here full localization does not occur. This means we can limit our study of the sigma model of class D only to the connected component of the manifold $\text{SO}(2n) / \text{U}(n)$ in the compact sector.

The WZW term in the action \eqref{sigma_model_action} is written in terms of the matrix $T$ rather than $Q$. In order to ensure that the action depends only on the $Q$-matrix configuration, we should apply some additional constraint on the choice of $T$. First, let us point out that the sigma model is well defined only for closed trajectories in terms of $Q$. This is consistent with the fact that the theory is derived for edge modes of a 2D topological superconductor, hence they are automatically closed. Second, the manifold of the sigma model of class D is simply connected (since we have already ruled out a possibility of jumps between two disjoint parts of the manifold). This means that any closed trajectory in terms of $Q$ can be continuously deformed into a single point. We will require that the corresponding trajectory in terms of $T$ is also topologically trivial and can be shrunk into a single point within the group $G$. Note that the group $G$, unlike the coset space $G/K$, is not simply connected. Hence this condition on $T$ is not trivial. The derivation of the action (\ref{sigma_model_action}) in Appendix \ref{appsigma} explicitly employs this topological property and demonstrates the invariance of the action with respect to reparametrizations of the field trajectory in terms of $T$.

\subsection{Cartan-Efetov parametrization}
\label{sec:Cartan-Efetov}

All physical observables of a disordered system can be inferred from the partition function of the nonlinear sigma model. For a system of 1D geometry this partition function is given by the path integral with the sigma-model action. The symmetry of the sigma-model manifold always allows us to bring the initial point of any trajectory to $\Lambda$. Hence we consider only path integrals for trajectories starting at $Q = \Lambda$:
\be
\label{partition_function}
  Z[Q]
  =
  \int_{Q(0) = \Lambda}^{Q(L) = Q} \mathcal{D}[Q(x)] e^{-S[Q]}.
\ee
The end point of the path integral can be also brought to its simplest form by some rotation $Q \mapsto U^{-1} \Lambda U$ with $[\Lambda, U] = 0$. Such rotations with a constant matrix $U$ do not shift the initial point and also preserve the form of the action. To take advantage of this symmetry we will use the Cartan decomposition of the group $G$ and define Cartan-Efetov coordinates on the sigma-model manifold.

Any matrix $T \in G$ can be decomposed into the following product:
\begin{equation}
 \label{T_parametrization}
 T = U_1 e^{\check\theta/2} U.
\end{equation}
Both matrices $U_1$ and $U$ commute with $\Lambda$ and thus belong to the subgroup $K \subset G$, while $\check\theta$ is taken from the maximal abelian subalgebra of group $G$ whose generators anticommute with $\Lambda$: $\{\Lambda, \check\theta\} = 0$. This representation is known as the Cartan decomposition of $T$ and $\check\theta$ is an element of the Cartan subalgebra. With such a form of the matrix $T$, we have
\be
 Q = U^{-1} \Lambda e^{\check\theta} U.
\ee
The matrix $Q$ is independent of $U_1$ and belongs to the coset space $G/K$ as was discussed above. We see that by a certain $U$ rotation it is always possible to bring the final point of our path integral (\ref{partition_function}) to the form $Q = \Lambda e^{\check\theta}$. Hence the partition function depends on the Cartan angles $\check\theta$ only.

Since the Cartan subalgebra is abelian it is always possible to choose a representation that brings $\check\theta$ to explicitly diagonal form [see Eq.\ (\ref{diag_theta})]. In class D with $n = 2$ replicas there are two Cartan angles in the non-compact boson sector and only one angle in the compact fermion sector. We will denote these angles as $\thone$, $\thtwo$, and $\thF$. Certain rotations $U$ can lead to permutations of the diagonal elements of the matrix $\check\theta$. Effectively this may lead to changing of the sign of any individual angle $\theta_i$ or to the interchange $\thone \leftrightarrow \thtwo$. To remove the ensuing uncertainty in the definition of $\check\theta$, we will assume
\begin{equation}
 \thone > \thtwo > 0, \qquad 0 < \thF < \pi.
 \label{domain}
\end{equation}

The partition function \eqref{partition_function} depends only on the values of Cartan angles at the final point of all trajectories. Once the partition function is known we can readily express first three moments of conductance by the following derivatives \cite{classD_first}:
\begin{subequations}
\label{G_through_Z}
\begin{gather}
\label{G1_through_Z}
	\left\langle g \right\rangle = -4 \left.\frac{\partial^{2}Z(\theta_i)}{ \partial \thone^2}\right|_{0},
\\{}
\label{G2_through_Z}
	\left\langle g^2 \right\rangle = 16 \left.\frac{\partial^{4}Z(\theta_i)}{ \partial \thone^2 \partial \thtwo^2}\right|_{0} ,
\\{}
\label{G3_through_Z}
	\left\langle g^3 \right\rangle = -32 \left.\frac{\partial^{6}Z(\theta_i)}{ \partial \thone^2 \partial \thtwo^2 \partial \thF^2}\right|_{0} .
\end{gather}
\end{subequations}
Here every derivative is taken at the point $\check\theta = 0$.

Strictly speaking, in the evaluation of the partition function \eqref{partition_function} we consider trajectories $Q(x)$ starting at $\Lambda$ and ending at $Q$ with certain non-zero values of the Cartan angles $\check\theta$. Such trajectories are not closed and the WZW term in the action is not well-defined for them as explained above. This issue can be resolved in the following way. We augment each trajectory going from $\Lambda$ to $Q$ with an additional segment that goes back from $Q$ to $\Lambda$ and makes the trajectory closed. The simplest choice is to take the shortest closing path along a geodesic line on the sigma-model manifold. We only need to add the value of the WZW term along this additional segment to the overall action. More detailed discussion of this closing procedure can be found in Ref.\ \onlinecite{KSO2016}. In the subsequent analysis, we will use a special gauge choice [see Eq.\ (\ref{T_gauge_choice}) below], that renders the contribution of the closing geodesic segment of every trajectory identically zero. This will allow us to disregard the contribution of outer edges and focus on the transport along inner interface only.

\begin{table}
\caption{Root system for the sigma-model manifold of class D with $n = 2$ replicas: positive roots ($\alpha$), their multiplicities ($m_\alpha$) and corresponding root vectors ($Z_{\alpha(,i)}$). The matrix $\Xi_{ij}$ has 1 at the position $(i,j)$ and $0$ elsewhere [in the basis with the diagonal matrix $\hat\theta$ and the charge conjugation matrix $C$ given by Eqs.\ (\ref{diag_theta}) and (\ref{C}), respectively].}
	\label{T:roots}
	\begin{ruledtabular}
	\begin{tabular}{ccccc}
\multicolumn{2}{c}{Bosonic ($m_\alpha=1$)} &
\multicolumn{3}{c}{Fermionic ($m_\alpha=-2$)} \\
\hline
$\alpha$ & $Z_{\alpha}$ &
$\alpha$ & $Z_{\alpha,1}$ & $Z_{\alpha,2}$ \\
\hline
$2 \thone$ & $\Xi_{18}$ &
$\thone + i \thF$ & $\Xi_{15}-\Xi_{38}$ & $\Xi_{16}+\Xi_{48}$ \\
$2 \thtwo$ & $\Xi_{27}$ &
$\thone - i \thF$ & $\Xi_{13}-\Xi_{48}$ & $\Xi_{14}+\Xi_{68}$ \\
$2 \thF$   & $\Xi_{36} + \Xi_{45}$ &
$\thtwo + i \thF$ & $\Xi_{25}-\Xi_{37}$ & $\Xi_{26}+\Xi_{47}$ \\
$\thone + \thtwo$ & $\Xi_{17} - \Xi_{28}$ &
$\thtwo - i \thF$ & $\Xi_{23} - \Xi_{57}$ & $\Xi_{24} + \Xi_{67} $ \\
$\thone - \thtwo$ & $\Xi_{12} - \Xi_{78}$
&&& \\
	\end{tabular}
	\end{ruledtabular}
\end{table}

The structure of the symmetric supermanifold of the sigma model can be further detailed by specifying the corresponding roots and root vectors\cite{Helgason}. A general matrix $T$ in the definition (\ref{Q_via_Lambda_and_T}) can be chosen in the exponential form $T = e^W$, where $W = - \overline W$. Matrices $W$ obey a linear constraints and thus span a certain linear space (Lie algebra of the group $G$). We construct a basis in this space by defining root vectors $Z_\alpha$. They are simultaneous eigenvectors of the adjoint action of all elements of the Cartan subalgebra: $[\check\theta, Z_\alpha] = \alpha(\check\theta) Z_\alpha$. Indeed, since Cartan subalgebra is abelian, it is always possible to simultaneously diagonalize all the commutators $[\check\theta, \cdot]$ and find the basis of root vectors $Z_\alpha$. The corresponding eigenvalue $\alpha(\check\theta)$ is a linear function of all the elements of $\check\theta$. It is called a root and can be viewed as a vector from the conjugate space to the Cartan subalgebra. We will include in the basis $\{Z_\alpha\}$ only nonzero root vectors, that is we exclude the elements of Cartan subalgebra from this set. If a certain simultaneous eigenvalue $\alpha(\check\theta)$ is degenerate and has more than one corresponding root vector $Z_\alpha$, we say that the multiplicity $m_\alpha$ of this root is greater than one. In a supersymmetric space there are bosonic and fermionic roots. Fermionic roots have Grassmann-valued root vectors and their degeneracy is always at least $2$. However, the corresponding multiplicity should be taken with a negative sign \cite{MMZ}: $m_\alpha = -2$.

Roots obey a set of remarkable geometrical properties that allows to fully classify all possible symmetric spaces by their root systems\cite{Helgason}. However, this goes beyond the scope of our paper. In the subsequent analysis we will only need the following simple property: if $\alpha(\check\theta)$ is a root, $-\alpha(\check\theta)$ is also a root. We can separate the full set of roots into two halves by drawing a hyperplane in the space of roots (conjugate to the Cartan subalgebra). All the roots that point to one side of this hyperplane are called positive roots; their set will be denoted $R^+$. The roots pointing in the other halfspace are called negative. For every positive root $\alpha$ there is always a corresponding negative root $-\alpha$.

It is always possible to represent matrices $t$ that generate the rotation $T$ such that the abelian Cartan subalgebra corresponds to diagonal matrices. Furthermore, it is also possible to bring the matrices representing all positive root vectors to explicitly upper triangular form. Then negative root vectors will be given by lower triangular matrices. We list roots and root vectors for the sigma-model manifold of symmetry class D with $n = 2$ replicas in Table \ref{T:roots}.

The knowledge of the root system gives us one important piece of information that will be extensively used below. Namely, it allows us to find the Jacobian of the Cartan parametrization \cite{Helgason}:
\begin{multline}
\label{Jacobian}
 J(\theta_i) =  \prod_{\alpha \in R^+} \left[ \sinh \frac{\alpha(\check\theta)}{2} \right]^{m_\alpha} \\
  = \frac{(\cosh \thone - \cosh \thtwo) \sinh \thone \sinh \thtwo \sin \thF}
	{(\cosh \thone - \cos \thF)^2 (\cosh \thtwo - \cos \thF)^2}.
\end{multline}
This Jacobian is the square root of the superdeterminant of the metric tensor of the sigma-model manifold and hence defines the invariant integration measure in Cartan angles $\check\theta$.

\subsection{Transfer-matrix approach}

Evaluation of the partition function \eqref{partition_function} can be accomplished within the transfer-matrix technique \cite{EL1983}. It amounts to mapping the path integral to the evolution operator of the corresponding quantum Hamiltonian in the imaginary time $x$. We can thus equate the partition function to the resulting wave function after the evolution:
\be
\label{Z_eq_psi}
	Z[\check\theta] = \Psi(Q = \Lambda e^{\check\theta}, x=L).
\ee
The result of such an evolution is usually termed \emph{the heat kernel}. Changing of $\Psi$ with $x$ is governed by the Schr{\"o}dinger equation with the Hamiltonian derived from the action (\ref{sigma_model_action}),
\be
\label{Schroedinger_equation}
	\frac{\xi}{2} \frac{\partial \Psi (Q, x)}{\partial x} = - \hat{H} \Psi(Q, x).
\ee
The initial condition is $\Psi(Q, x=0) = \delta(Q, \Lambda)$, with the delta function that fixes $Q = \Lambda$.

The heat kernel can then be expressed as the spectral decomposition over the eigenfunctions $\phi_\nu$ of $\hat{H}$:
\be
\label{heat_kernel}
	\Psi(Q,x) = \sum_\nu \mu_\nu \phi_\nu(Q) e^{-2 \epsilon_\nu x / \xi},
\ee
where $\epsilon_\nu$ are the eigenvalues and $\mu_\nu$ is a proper measure.

In a given coordinate system $X_\alpha$ on the sigma-model manifold, the transfer-matrix Hamiltonian for the action \eqref{sigma_model_action} has the form \cite{Eslam_thesis}:
\be
\label{Hamiltonian_general}
	\hat{H}  = - \frac{1}{J} \left(\partial_\alpha - A_\alpha \right) J g^{\alpha \beta} \left(\partial_\beta - A_\beta \right).
\ee
Here $g_{\alpha \beta}$ is the metric tensor related to the length element as \cite{Efetov-book, Eslam_thesis}
\be
  -\frac{1}{2} \str dQ^2 = g_{\alpha \beta} dX^{\alpha} dX^{\beta}.
\ee
The factor $J = \sqrt{\sdet g}$ is the Jacobian [for Cartan coordinates it is given in Eq.\ (\ref{Jacobian})]. A new ingredient in the Hamiltonian \eqref{Hamiltonian_general} compared to the $m = 0$ case is the appearance of a vector potential $A_\alpha$ generated by the WZW term:
\be\label{vector_potential}
	A_\alpha = \frac{m}{4} \str T^{-1} \Lambda \partial_\alpha T.
\ee
As was already discussed earlier, this vector potential is written in terms of $T$ rather than $Q$, and hence is not gauge invariant. Fixing the gauge means assigning a particular unique matrix $T$ to every value of $Q$. This will allow to express $A_\alpha$ as a function of $Q$.

We will use the gauge
\be
\label{T_gauge_choice}
	T = U^{-1} e^{\check{\theta} / 2} U.
\ee
It means that we fix the matrix $U_1 = U^{-1}$ in the Cartan decomposition (\ref{T_parametrization}) of $T$. Substituting Eq.\ (\ref{T_gauge_choice}) into Eq.\ (\ref{vector_potential}), it is easy to see that the vector potential depends on the Cartan angles $\hat\theta$ only. This property will greatly simplify further analysis. Indeed, initial condition of the evolution $Q = \Lambda$ is invariant under rotations $Q \mapsto U^{-1} Q U$ with any matrix $U$ that commutes with $\Lambda$. The same property is also ensured for the Hamiltonian (\ref{Hamiltonian_general}) in the chosen gauge (\ref{T_gauge_choice}). Hence we can limit our consideration to the ``radial'' eigenfunctions $\phi_\nu$ that depend only on $\check\theta$.

This situation is similar to solving quantum evolution problem for a charged particle on a 2D plane subject to a uniform magnetic field. If one chooses circular gauge with the origin that coincides with the initial position of the particle the whole evolution operator at all times will depend only on the radial coordinate but not on the polar angle.

We can write explicitly the radial Hamiltonian that involves only the Cartan angles $\check\theta$:
\be
	\label{Hamiltonian_radial}
	H = - \Delta + W(\check\theta),
\ee
where the radial Laplace operator is
\be
\label{radial_laplacian}
	\Delta = \frac{1}{J} \left( \frac{\partial}{\partial \thone} J \frac{\partial}{\partial \thone}
	+ \frac{\partial}{\partial \thtwo} J \frac{\partial}{\partial \thtwo}
	+ \frac{1}{2} \frac{\partial}{\partial \thF} J \frac{\partial}{\partial \thF}
	\right)
\ee
and the Jacobian is given by Eq.\ (\ref{Jacobian}). The WZW vector potential $A_\alpha$ enters the radial Hamiltonian \eqref{Hamiltonian_radial} through the potential term $W(\check\theta) = - A_\alpha A^\alpha$. For the sigma model of class D with $n = 2$ replicas it reads
\be
\label{potential_explicit}
	W(\check\theta) = \frac{m^2}{16} \left( \tanh^2 \frac{\thone}{2} + \tanh^2 \frac{\thtwo}{2} + 2 \tan^2 \frac{\thF}{2} \right) .
\ee
In the next section we will explicitly construct eigenfunctions of the Hamiltonian (\ref{Hamiltonian_radial}).

In the limit of large noncompact angles $\thone$ and $\thtwo$, the transfer-matrix Hamiltonian (\ref{Hamiltonian_radial}) greatly simplifies. Assuming $\thone \gg \thtwo \gg 1$ [note how the domain (\ref{domain}) restricts possible values of $\check\theta$], we reduce it to the form
\begin{multline}
 H
  \approx -\frac{\partial^2}{\partial\thone^2} - \frac{\partial^2}{\partial\thtwo^2} + \frac{\partial}{\partial\thtwo} \\
     + \frac{1}{2} \left( -\frac{\partial^2}{\partial\thF^2} - \cot\thF \frac{\partial}{\partial\thF} + \frac{m^2}{4 \cos^2(\thF/2)} \right).
\end{multline}
Since all the variables are decoupled, we can easily find the limiting form of an eigenfunction:
\begin{equation}
 \phi_{\bm{q}}(\check\theta)
  \approx e^{i q_1 \thone + (i q_2 + 1/2) \thtwo} \cos^m(\thF/2) P^{(0,m)}_{l - m/2}(\cos\thF) ,
 \label{asympP}
\end{equation}
with $P$ being the Jacobi polynomial. The corresponding eigenvalue is
\begin{equation}
 \epsilon_{\bm{q}}
  = q_1^2 + q_2^2 + \frac{l(l + 1)}{2} + \frac{1}{4}.
 \label{eigenvalue}
\end{equation}
The asymtotic plane waves in the noncompact angles are parametrized by momenta $q_{1,2}$, while the dependence on the compact angle $\thF$ is given by the Jacobi polynomial whose index is quantified by $l$. For brevity, we use the single index $\bm{q}$ to denote all three components of momentum.

In order to ensure that $\sqrt{J} \phi$ does not exponentially grow at large values of $\check\theta$, we assume $q_{1,2}$ real. At the same time, the eigenfunction has a well-defined single value on the segment $0 < \thF < \pi$ provided the Jacobi polynomial index $l - m/2$ is a non-negative integer. From this analysis of the asymptotic behavior we conclude that the presence of the WZW term in the action and hence the potential term in the transfer-matrix Hamiltonian (\ref{Hamiltonian_radial}) do not alter the spectrum of eigenvalues \eqref{eigenvalue} in its general form. However, in the following we will show that there are additional families of eigenfunctions, which have $m$-dependent eigenvalues.

\subsection{Iwasawa parametrization}
\label{sec:Iwasawa_trick}

Eigenfunctions of the radial Laplace operator \eqref{Hamiltonian_radial} can be efficiently constructed by considering the full Laplace operator in Iwasawa coordinates \cite{Helgason, Zirnbauer1992, MMZ}. This representation is based on the following decomposition of the matrix $T$:
\be
\label{T_Iwasawa}
 T = V e^{\check a/2} N.
\ee
Here $V$ is a matrix that commutes with $\Lambda$, $\check a$ is an element of the Cartan subalgebra parametrized by three angles $\aone$, $\atwo$, and $\aF$, and $N = e^n$ is an exponential of a nilpotent matrix $n$ that represents a linear combination of positive root vectors.

It is known that in the $m = 0$ case this parametrization has the following important property. The radial Laplace operator in Iwasawa coordinates [part of the full Laplace operator (\ref{Hamiltonian_general}) that involves only $\check a$] does not explicitly depend on $\check a$. Instead, it contains only derivatives with respect to $\check a$ and hence simple plane waves $\exp{\left( i \sum_i p_i a_i \right)}$ are the radial eigenfunctions.

It turns out \cite{Eslam_thesis} that this property does persist in the presence of the WZW term when the proper gauge choice is made. Namely, one should assume $V = 1$ in the decomposition (\ref{T_Iwasawa}). Then the $\check a$-dependent part of the Hamiltonian takes the following simple form:
\be
\label{Iwasawa_Laplacian}
  \hat H
  =
  -\left(
    \frac{\partial^2}{\partial a_\text{B1}^2} + \frac{\partial^2}{\partial a_\text{B2}^2} + \frac{1}{2} \frac{\partial^2}{\partial a_\text{F}^2}
    - \frac{\partial}{\partial a_\text{B2}} + \frac{i}{2} \frac{\partial}{\partial a_\text{F}}
  \right).
\ee
The radial Laplace operator in the Iwasawa coordinates is not only independent of $\check a$ but also does not contain any potential whatsoever!

Eigenfunctions of the radial Hamiltonian (\ref{Iwasawa_Laplacian}) in Iwasawa coordinates are plane waves
\be
\label{plane_wave}
 \phi_{\bm{q}}(\check a)
  = \exp\left[ i q_1 a_\text{B1} + \left( i q_2 + {1}/{2} \right) a_\text{B2} + i l a_\text{F} \right].
\ee
Below we will also use a short-hand notation for this linear combination: $\phi_{\bm{q}}(\check a) = e^{i \bm{q} \cdot a}$. The corresponding eigenvalue for this plane wave is given by Eq.\ (\ref{eigenvalue}).

We have thus established a set of radial eigenfunctions in the Iwasawa coordinates. In order to calculate the heat kernel (\ref{heat_kernel}), we will need radial eigenfunctions in the Cartan-Efetov coordinates. Let us first relate the two sets of coordinates to each other:
\be
\label{gauge_change}
 T
  = V e^{\hat a/2} N
  = U^{-1} e^{\check\theta/2} U.
\ee
This matrix identity allows us to express $\check a$ and $V$ as functions of $\check\theta$ and $U$. In Ref.\ \onlinecite{classD_first}, we have discussed in detail how to resolve these equations for $\check a$. Solution for the matrix $V(\check\theta, U)$ is explained in Appendix \ref{app:gauge}. Let us note that we have chosen here $T$ in the proper radial gauge (\ref{T_gauge_choice}) for the Cartan parametrization. However, this does not correspond to the proper gauge in Iwasawa coordinates since $V \neq 1$. We can account for this new gauge choice in the Iwasawa coordinates by multiplying the plane wave eigenfunction (\ref{plane_wave}) with the proper phase factor [cf.\ Eq.\ (\ref{gauge_action})]:
\begin{equation}
\label{gauged_wavefunction}
 \phi_{\bm{q}}(\check\theta, U)
  =  e^{(m/4) \str [\Lambda \ln V(\check\theta, U)]} e^{i \bm{q} \cdot a (\check\theta, U)}.
\end{equation}

Further steps in the derivation of radial wave functions follow Refs.\ \onlinecite{Eslam_thesis, classD_first}. By construction, the new function (\ref{gauged_wavefunction}) is automatically an eigenfunction of the full Hamiltonian (\ref{Hamiltonian_general}) in Cartan coordinates $\check\theta$ and $U$ with the gauge (\ref{T_gauge_choice}). However, this eigenfunction does explicitly depend on $U$. In order to obtain the corresponding radial eigenfunction that depends on $\check\theta$ only, one has to perform \emph{isotropization} of \eqref{gauged_wavefunction} over the group $K$ of all possible matrices $U$:
\be
\label{isotropization}
  \phi_{\bm{q}} (\check\theta)
  = \left< e^{(m/4) \str [\Lambda \ln V(\check\theta, U)]} e^{i \bm{q} \cdot a (\check\theta, U)} \right>_{U \in K}.
\ee

Usually (for noncompact symmetric spaces) isotropization \eqref{isotropization} is performed by integration over $K$ group. However, as explained in detail in Ref.\ \onlinecite{classD_first}, the peculiarity of the supersymmetric model is that a na\"ive integration over the whole $K$ group leads to the loss of some eigenfunctions. This happens because for certain values of momentum vector $\bm{q}$ the integrand lacks some of the Grassmann variables and the whole integral vanishes. A correct strategy to obtain these extra eigenfunctions is to integrate only over required Grassmann variables, which can be achieved by a special parametrization of the $U$ matrix.

Radial eigenfunctions defined by Eq.\ (\ref{isotropization}) are symmetric under changing sign of any of $q_{1,2}$, mapping $l \mapsto -l - 1$, or interchanging $q_1 \leftrightarrow q_2$. This is a consequence of the Weyl group symmetry of the root system. Therefore we can limit the space of eigenfunctions to the domain $q_1 > q_2 >0$ and to the non-negative values of $l$ only.

\section{Structure of the eigenbasis}
\label{sec:eigenbasis}

In the $m = 0$ case \cite{classD_first}, we found that the radial eigenbasis of the Laplace-Beltrami operator on the $n = 2$ sigma-model supermanifold has a hierarchical structure and consists of three families: (i) the zero mode (just unity for $m = 0$), (ii) one-parameter family, and (iii) the most generic three-parameter family. While the latter is routinely obtained via the Iwasawa trick, identifying one-parameter family and the zero mode following the same approach is a non-trivial task. It requires modification of the standard procedure to avoid nullification of the integral \eqref{isotropization} due to the lack of a complete set of Grassmann variables in each term of the integrand.

Below we apply Eq.\ (\ref{isotropization}) to obtain the eigenfunctions of the transfer-matrix Hamiltonian \eqref{Hamiltonian_radial} in the $m\neq0$ case. We compute all three families of radial eigenfunctions for arbitrary value of $m$. While it is in general impossible to get closed analytic expressions for the eigenfunctions, we will identify their large- and small-$\theta$ asymptotics. This will be sufficient to properly normalize the eigenfunctions, find the spectral decomposition of the heat kernel (\ref{heat_kernel}), and compute transport moments (\ref{G_through_Z}).

\subsection{Zero mode}
It can be checked by a direct computation that the following function is the zero mode of the Hamiltonian~\eqref{Hamiltonian_radial}:
\be
\label{zero_mode}
	\Psi_0 = \frac{\cos^m(\thF/2)}{\cosh^{m/2}(\thone/2) \cosh^{m/2}(\thtwo/2)}.
\ee
This function can be derived from Eq.\ (\ref{isotropization}) by setting $q_1 = im/4$, $q_2 = im/4 + i/2$, and $l = m/2$. [With these values of momenta, the wave function does not depend on the Iwasawa angles $a_i$, as can be seen from Eq.\ (\ref{isotropization2}).] The corresponding eigenvalue is zero according to Eq.\ (\ref{eigenvalue}). In the $m=0$ case this expression equals unity, which is known to be the zero mode in the supersymmetric theories without a WZW term. For a nonzero $m$, the eigenfunction automatically nullifies at the ``south pole'' $\thF = \pi$, the property discussed in detail in Ref.\ \onlinecite{KSO2016}.

\subsection{One-parameter family}
\label{sec:eigenfunctions_1P}

The one-parameter family of eigenfunctions appears when only the first Iwasawa angle $\aone$ is present in Eq.\ (\ref{isotropization}) and corresponds to $q_2 = im/4 + i/2$ and $l = m/2$, while $q_1$ is an arbitrary real number. The numerator of Eq.\ (\ref{isotropization2}) indeed contains only $\aone$ in this case. The eigenvalue is
\begin{equation}
 \epsilon_{q_1} = q_1^2 + \frac{m^2}{16} ,
 \label{eps1}
\end{equation}
as follows from Eq.\ (\ref{eigenvalue}). These eigenfunctions are in one-to-one correspondence with the eigenfunctions of the transfer-matrix Hamiltonian for the sigma model with $n=1$ replica. We can establish this relation by inspecting the Hamiltonian (\ref{Hamiltonian_radial}) in the vicinity of the ``bosonic'' line $\thtwo = \thF = 0$. [Note that a similar line $\thone = \thF = 0$ is excluded from the domain of definition (\ref{domain}).] Expand an eigenfunction as
\begin{equation}
 \phi
  \approx \psi(\thone) + u(\thone) \thtwo^2 + v(\thone) \thF^2.
\end{equation}
Acting with the Hamiltonian (\ref{Hamiltonian_radial}) on this function and expanding the result in small $\thtwo$ and $\thF$, we obtain
\begin{multline}
 H\phi
  = \left(-\frac{\partial^2}{\partial \thone^2} + \frac{1}{\sinh\thone} \frac{\partial}{\partial\thone} + \frac{m^2}{16} \tanh^2\frac{\thone}{2} \right) \psi \\
    + \frac{\thtwo^2 - \thF^2}{\thtwo^2 + \thF^2} ( 4 u - 2 v ).
 \label{Laplace1}
\end{multline}
From this expression we see that a unique limiting value of the function at the bosonic line $\thtwo = \thF = 0$ exists only provided $v = 2u$. At the same time, the equation for $\psi(\thone)$ acquires a form of the transfer-matrix Hamiltonian of class D with only one replica acting on a function of only one Cartan angle $\thone$. Hence we can readily identify the eigenfunction in this limit
\begin{multline}
 \phi_{q_1}(\thone)
  = -\left( q_1^2 + \frac{m^2}{16} \right) \frac{2\sinh^2(\thone/2)}{\cosh^{m/2}(\thone/2)} \\
    \times F\left( 1 - \frac{m}{4} + iq_1,\, 1 - \frac{m}{4} - iq_1;\, 2;\, -\sinh^2\frac{\thone}{2} \right).
 \label{psiq}
\end{multline}

Asymptotic expansion at large values of $\thone$ yields a plane wave of the following form:
\begin{gather}
 \phi_{q_1}(\thone \gg 1)
  = c_{q_1} e^{i q_1 \thone} + c_{-q_1} e^{-i q_1 \thone}, \label{psi1c} \\
 c_{q_1}
  = \frac{2^{1-2iq_1}\Gamma(2iq_1)}{\Gamma(iq_1+m/4)\Gamma(iq_1-m/4)}.
 \label{c1}
\end{gather}
The amplitude $c_{q_1}$ that appears here is called Harish-Chandra $c$-function. It can be derived directly from Eq.\ (\ref{isotropization}) without referring to the exact function (\ref{psiq}). Knowledge of $c_{q_1}$ is sufficient to normalize the eigenfunction as we will show later.

We can also use Eq.\ \eqref{isotropization} to find the eigenfunction at the line $\thone = \thtwo = 0$ for arbitrary $\thF$:
\be
  \label{psi1P_fermionic_line}
  \phi_{q_1}(\thF)
  = -4 (q_1^2+m^2/16) \cos^m(\thF/2)\sin^2(\thF/2).
\ee
This expression allows us \cite{classD_first} to recover the full small-$\theta$ expansion of the eigenfunctions to be used in evaluating transport properties according to Eqs.\ \eqref{G_through_Z}.

The one-parameter family of eigenfunctions is parametrized by real values of the first momentum component $q_1$. The zero mode (\ref{zero_mode}) can be also restored from the one-parameter family if we set momentum to the imaginary value $q_1 = im/4$. For sufficiently large values of $m$, a finite set of discrete eigenfunctions appears with $q_1 = i(m/4 - 1)$, $i(m/4 - 2)$, $i(m/4 - 3)$ etc.\ as long as these imaginary values are positive. These functions exponentially decay in the limit $\thone \gg 1$ as can be seen from Eq.\ (\ref{c1}). They correspond to eigenstates localized in the potential $W$, cf.\ Eq.\ (\ref{Hamiltonian_general}). We will discuss these functions in more detail later.

\subsection{Three-parameter family}
\label{sec:eigenfunctions_3P}

Generic radial eigenfunctions of the Hamiltonian (\ref{Hamiltonian_radial}) are parametrized by three components of momentum and are given by the general expression (\ref{isotropization}). The values of $q_1$ and $q_2$ are arbitrary real numbers and $l - m/2$ must be an integer. The corresponding eigenvalue is given by Eq.\ (\ref{eigenvalue}).

It is not feasible to calculate the full integral over $U$ in Eq.\ (\ref{isotropization}) at arbitrary $\check\theta$. However, we can extract a limiting form of the eigenfunction at large values of $\check\theta$. In order to do so, we will temporarily assume that momenta $q_1$ and $q_2$ have a small negative imaginary part
\begin{equation}
 \operatorname{Im} q_1 < \operatorname{Im} q_2 < 0.
 \label{Imq}
\end{equation}
Then the limiting plane wave (\ref{asympP}) exhibits some extra exponential growth and dominates at large $\thone \gg \thtwo \gg 1$ over other terms. Calculation of the integral (\ref{isotropization}) under this assumptions yields the Harish-Chandra $c$-function which is just the prefactor for the asymptotic expression (\ref{asympP}):
\begin{gather}
 \phi_{\bm{q}}
  \approx c_{q_{1,2}}\, e^{i q_1 \thone + (i q_2 + 1/2) \thtwo} \cos^m(\thF/2) P_{l - m/2}^{(0,m)}(\cos\thF), \label{asymp_psi} \\
 c_{q_{1,2}}
  = \frac{
      (l + 2 i q_1) (l + 2 i q_2) (1 + l - 2 i q_1) (1 + l - 2 i q_2)
    }{
      \pi C_{q_1}^{(m)} C_{q_2}^{(m)} C_{q_1 + q_2}^{(0)} C_{q_1 - q_2}^{(0)}
    },  \label{c3}\\
 C_q^{(m)}
  = \frac{\Gamma\left(\frac{1}{2} + i q + \frac{m}{4} \right) \Gamma\left(\frac{1}{2} + i q - \frac{m}{4} \right)}{\sqrt{2\pi}\, 2^{-2 i q} \Gamma(2 i q)} .
\end{gather}
For brevity we are again using a single index $\bm{q}$ to denote all three components of momentum $q_1$, $q_2$, and $l$.

Now we remove the above assumption (\ref{Imq}) and consider real values of $q_{1,2}$. In this case the plane wave (\ref{asymp_psi}) is no longer the only dominant term for large $\check\theta$. A full eigenfunction must be symmetric with respect to the Weyl group acting on momentum. That is, the function should be invariant under changing sign of $q_1$ and/or $q_2$ as well as under interchanging $q_1 \leftrightarrow q_2$. Thus we can restore the full asymptotics of the eigenfunction by averaging Eq.\ (\ref{asymp_psi}) over the Weyl group
\begin{equation}
 \phi_{\bm{q}}
  \approx \cos^m(\thF/2) P_{l - m/2}^{(0,m)}(\cos\thF) e^{\thtwo/2} \sum_{w \in W} c_{w q} e^{i \langle w q | \theta \rangle}.
 \label{asymp_psi_sym}
\end{equation}
Here we use the following Dirac notations in the noncompact variables: $\langle q | \theta \rangle = q_1 \thone + q_2 \thtwo$ and denote by $wq$ the action of the Weyl group element $w$ in the noncompact sector on the momentum $q$. Weyl group in the compact sector contains the only symmetry operation $l \mapsto -1 - l$. This symmetry is fulfilled by Eq.\ (\ref{asymp_psi_sym}) automatically due to the properties of Jacobi polynomials.

Besides the large-$\theta$ asymptotics, Eq.\ \eqref{isotropization} allows us to compute the eigenfunction explicitly at the line $\thone = \thtwo = 0$:
\begin{multline}
 \phi_{\bm{q}}(\thF)
   = \cos^m(\thF/2) \sin^4(\thF/2) \\
     \times \frac{(4q_1^2 + l^2)(4q_2^2 + l^2)(4l^2 - m^2)}{2l + 1} \\
     \times F\left( 1 - l + \frac{m}{2},  1 + l + \frac{m}{2}; 2; \sin^2\frac{\thF}{2} \right) \\
     +\bigl\{ l \mapsto -l-1 \bigr\}.
 \label{psi3P_fermionic_line}
\end{multline}
We will use this result later to infer \cite{classD_first} the small-$\theta$ expansion required in Eqs.\ \eqref{G_through_Z}.

For large enough values of $m$ some extra eigenfunctions appear with either one or both $q_{1,2}$ taking discrete imaginary values. We will discuss these cases in the next section.

\section{Heat kernel construction}
\label{sec:heat_kernel}

Now, once all the eigenstates of the transfer-matrix Hamiltonian are identified, we are in a position to write the heat kernel as a sum over these eigenstates. According to Eq.\ (\ref{heat_kernel}), we have
\begin{equation}
 \Psi
  = \sum_\nu \mu_\nu \phi_\nu e^{-2 \epsilon_\nu x / \xi}
  = \Psi_0 + \Psi^{(1)} + \Psi^{(3)}.
 \label{Psi}
\end{equation}
Here the index $\nu$ enumerates all the eigenstates of the transfer-matrix Hamiltonian, $\epsilon_\nu$ are the corresponding eigenvalues, and $\mu_\nu$ is the proper weight factor yet to be established. Since the Hamiltonian has both discrete and continuous branches of eigenstates, summation over $\nu$ implies either a sum or an integral where appropriate.

As discussed in Sec.~\ref{sec:eigenbasis}, there are in total three different families of eigenstates: the zero mode $\Psi_0$, one-parameter family as in the model with one replica, and generic three-parameter family. We have indicated these contributions separately in the right-hand side of Eq.\ (\ref{Psi}). The zero mode (\ref{zero_mode}) enters the expansion (\ref{Psi}) with the unit weight. This is the only eigenfunction that is nonzero at the origin $\check\theta = 0$ and its weight is fixed by the initial condition for the evolution equation (\ref{heat_kernel}). Two remaining parts of the heat kernel are considered below.

The main guiding principle of our construction is the following. We do know the correct expansion of the heat kernel in the case $m = 0$ from Ref.\ \onlinecite{classD_first}. The weights of the wave functions in the heat kernel expansion (\ref{Psi}) are expressed through the Harish-Chandra $c$-function describing the large-$\theta$ asymptotics. We assume that a similar relation holds in the case of non-zero $m$ with the functions (\ref{c1}) and (\ref{c3}). In order to account for all relevant discrete eigenstates of the Hamiltonian (\ref{Hamiltonian_radial}), which appear at non-zero $m$, we will construct the heat kernel by performing analytic continuation of the momentum integrals in Eq.\ (\ref{Psi}) in the parameter $m$ starting at $m = 0$.

\subsection{One-parameter family}

The weights for the one-parameter family of eigenfunctions (\ref{psiq}) can be deduced from the heat kernel on the ``bosonic'' line $\thtwo = \thF = 0$. This is equivalent to finding the heat kernel expansion for the one-replica model \cite{Eslam_thesis}. All the generic (three-parameter) eigenfunctions vanish on this line hence we can disregard them for now and write the heat kernel as
\begin{gather}
 \Psi(\thone)
  = \Psi_0 + \Psi^{(1)},
 \\
 \Psi^{(1)}
  = \int dq_1\, \mu_{q_1} \phi_{q_1} e^{-2 \epsilon_{q_1} x / \xi}.
 \label{Psi_fermi}
\end{gather}
Here the eigenvalue is given by Eq.\ (\ref{eps1}).

The weight $\mu_{q_1}$ can be inferred from the $\thone \gg 1$ limit (\ref{psi1c}) of the eigenfunction (in quantum mechanics, this approach is known as normalization of the wave functions of continuous spectrum by their asymptotics):
\begin{multline}
 \mu_{q_1}
  = \frac{1}{\pi c_{q_1} c_{-q_1}}
  = \frac{8 q_1}{16 q_1^2 + m^2} \\ \times \left[ \coth\left( \pi q_1 + \frac{i\pi m}{4} \right) + \coth\left(\pi q_1 - \frac{i \pi m}{4} \right) \right].
 \label{muq}
\end{multline}
This weight function is even in $q_1$ since the eigenfunctions with momenta $\pm q_1$ actually coincide. This is just a manifestation of the Weyl group symmetry for the model with one replica. Hence in the integral (\ref{Psi_fermi}) we are supposed to retain only positive $q_1$ in order to count each eigenfunction once. This provides a correct heat kernel expansion on the line $\thtwo = \thF = 0$ in the case $m = 0$. For nonzero $m$, we should take into account possible localized states that correspond to some discrete imaginary values of $q_1$, as was pointed out in the end of Sec.\ \ref{sec:eigenfunctions_1P}. We will achieve this by analytic continuation of the integral (\ref{Psi_fermi}) in the parameter $m$ starting at $m = 0$.

We begin with ``unfolding'' the Weyl group symmetry and introduce the modified weight factor
\begin{equation}
 \tilde\mu_{q_1}
  = \frac{8 q_1}{16 q_1^2 + m^2} \coth\left(\pi q_1 - \frac{i \pi m}{4} \right) ,
 \label{tildemuq}
\end{equation}
implying an integration over all real values of $q_1$ in Eq.\ (\ref{Psi_fermi}). The unfolded weight function (\ref{tildemuq}) has poles at the following imaginary values of $q_1$:
\begin{equation}
 q_1 = \frac{im}{4} + in, \qquad n \in \mathbb{Z}.
 \label{poles}
\end{equation}
[There is also a pair of poles at $q_1 = \pm im/4$ from the denominator of Eq.\ (\ref{tildemuq}) but these poles are canceled by the same factor in the numerator of the eigenfunction itself, cf.\ Eq. (\ref{psiq}).] Since we have ``unfolded'' the $q_1$ integral and retained only one cotangent factor in Eq.\ (\ref{tildemuq}), these poles move only upwards in the complex plane with increasing $m$. This fact is crucial for the analytic continuation of the integral (\ref{Psi_fermi}) in the parameter $m$. There is also one special pole at $q_1 = im/4$ corresponding to $n = 0$ in the series (\ref{poles}). This pole lies exactly on the integration contour when $m = 0$. However its residue is zero due to the factor $q_1$ in the numerator of Eq.\ (\ref{tildemuq}). If we deviate from $m = 0$ by assuming a small positive value of $m$, this pole acquires a finite residue but also shifts in the upper complex plane and lies now above the real axis.

To maintain analyticity of the heat kernel as a function of $m$ we first shift the integration contour in the lower complex plane of $q_1$ by $-i/2$ to stay away from the pole at $q_1 = im/4$ (we can as well shift by any other negative amount between $0$ and $-i$). Then with increasing $m$ we will simultaneously shift the integration contour upwards by $im/4$. This way we will avoid any possible crossing of the integration contour by the poles (\ref{poles}). With all these modifications, we can write the one-parameter part of the heat kernel as
\begin{equation}
 \Psi^{(1)}
  = \int_{-\infty - i/2 + im/4}^{\infty - i/2 + im/4} dq_1\, \tilde\mu_{q_1} \phi_{q_1} e^{-2 \epsilon_{q_1} x / \xi}.
\end{equation}

We have thus obtained an exact expression for the one-replica heat kernel in terms of the momentum integral with shifted integration contour. To extract the information about true eigenstates of the Hamiltonian we will now shift the contour back to the real axis collecting all the relevant pole contributions. After this backward shift we can also restore the Weyl group symmetry and ``fold'' the measure back to Eq.\ (\ref{muq}) keeping only positive real values of $q_1$. This transformation yields
\begin{multline}
 \Psi^{(1)}
  = \int_0^\infty dq_1\, \mu_{q_1} \phi_{q_1} e^{-2 \epsilon_{q_1} x / \xi} \\
    -\sum_{n = 1}^{\lfloor m/4 \rfloor} \frac{16 i q_1}{16 q_1^2 + m^2}\, \phi_{q_1} e^{-2 \epsilon_{q_1} x / \xi} \Bigr|_{q_1 = im/4-in}.
 \label{Psi1}
\end{multline}
We see that aside from the usual eigenfunctions with real positive $q_1$ there is also a finite set of discrete eigenstates corresponding to imaginary values of $q_1$. They are the eigenstates of the Hamiltonian (\ref{Hamiltonian_radial}) localized by the potential term $W$. The first such localized eigenfunction appears when $m$ exceeds $4$.

\subsection{Three-parameter family}

Let us now discuss the contribution $\Psi^{(3)}$ of generic eigenfunctions with all three non-trivial components of momentum to the heat kernel (\ref{Psi}).
As before, we can identify the weight of such eigenstates using the value of their $c$-function (\ref{c3}):
\begin{gather}
 \mu_{\bm{q}}
  = \frac{2l+1}{\pi^2 c_{q_{1,2}} c_{-q_{1,2}}}
  = \rho_{\bm{q}} T_{q_1}^{(m)} T_{q_2}^{(m)} T_{q_1 + q_2}^{(0)} T_{q_1 - q_2}^{(0)}, \label{mu3full} \\
 \rho_{\bm{q}}
  = \frac{(2l + 1) q_1 q_2 (q_1^2 - q_2^2)}{[4q_1^2 + l^2] [4q_2^2 + l^2] [4q_1^2 + (l + 1)^2] [4q_2^2 + (l + 1)^2]}, \\
 T_{q}^{(m)}
  = \tanh\left( \pi q + \frac{i \pi m}{4} \right) + \tanh\left( \pi q - \frac{i \pi m}{4} \right),
\end{gather}
where $\bm{q} = (q_1,q_2,l)$.

Contribution of the generic eigenfunctions to the heat kernel is
\begin{equation}
 \Psi^{(3)}
  = \sum_{l=m/2}^\infty \int dq_{1,2}\, \mu_{\bm{q}}\, \phi_{\bm{q}}\, e^{-2 \epsilon_{\bm{q}} x/\xi}.
\end{equation}
Here the eigenvalues $\epsilon_{\bm{q}}$ are given by Eq.\ (\ref{eigenvalue}). Summation over the compact momentum component $l$ starts with $m/2$ and proceeds in integer steps: $m/2$, $m/2 + 1$, $m/2 + 2$ etc. The domain of integration and contours for $q_{1,2}$ are yet to be established. In the case $m = 0$ this integral should be taken over the region $q_1 > q_2 > 0$ which is known as the Weyl chamber in momentum space. All other real values of $q_{1,2}$ can be mapped into this region by an element of the Weyl group.

To find appropriate integration limits for $q_{1,2}$ we first ``unfold'' the Weyl group symmetry and extend the integration to all real values of $q_{1,2}$. This is done in full analogy with the one-parameter eigenfunctions studied above. Using the symmetries of the measure, we can reduce the weight down to the following simple product of only two $\tanh$ factors:
\begin{equation}
 \tilde\mu_{\bm{q}}
  = 4 \rho_{\bm{q}} \tanh(\pi q_1 - \pi q_2) \tanh(\pi q_1 - i \pi m/4).
 \label{mu3}
\end{equation}
It is straightforward to check that symmetrization of $\tilde\mu_{\bm{q}}$ with respect to the Weyl group restores the full expression (\ref{mu3full}).

Consider first the poles of this unfolded measure as a function of $q_1$. We have a set of poles from the last $\tanh$ factor and also two pairs of poles due to the denominator of $\rho_\mathbf{q}$:
\begin{align}
 & q_1 = \frac{im}{4} + \frac{i}{2} + in, \hspace{5.7mm}
   \text{with $n \in \mathbb{Z}$;}
\label{poles21}
\\
 & q_1 = \pm \left(\frac{im}{4} + in\right), \quad
   \text{with $n = 0,1,2,3\ldots$}
\label{poles22}
\end{align}
In the second series of poles either $n = l - m/2$ or $n = l - m/2 + 1$ hence it is a non-negative integer. The first series of poles shifts only upwards with increasing $m$ so we can avoid crossing with these poles shifting the integration contour in $q_1$ by the same amount $im/4$. In the second set, poles with positive imaginary part are moving upwards while negative poles are moving downwards. Hence we do not have any chance of crossing them except for the lowest one $q_1 = im/4$. This pole exists only in the very first term of the sum (\ref{mu3}) with $l = m/2$. To avoid crossing this pole we will additionally shift our integration contour by an infinitely small amount $-i0$.

Possible poles in the variable $q_2$ come from the factor $\tanh(\pi q_1 - \pi q_2)$ in Eq.\ (\ref{mu3}). We can avoid these poles by shifting the contour for $q_2$ by the same amount as the contour for $q_1$. Then the difference $q_1 - q_2$ will always remain real and we never hit any singularity of the $\tanh$ function. The denominator of $\rho_q$ also provides a set of poles for $q_2$ similar to Eq.\ (\ref{poles22}). However these poles also stay away from the integration contour as long as we apply the same shift to $q_2$ as to $q_1$.

\begin{table*}
\caption{Coefficients $\chi_{q_1}^{(s)}$ and $\chi_{\bm{q}}^{(s)}$ in Eq.\ \eqref{result_general}, representing contributions of one- and three-parameter families to $\corr{g^s}$, correspondingly. The bottom line contains similar expressions for the (pseudo)Fano factor. In the second column, $\epsilon$ stands for the eigenvalue, as given by Eq.\ (\ref{eps1}).}
\begin{ruledtabular}
\begin{tabular}{ccc}
\rule[-8pt]{0pt}{20pt}
  & 
$\chi_{q_1}^{(s)}$ & $\chi_{\bm{q}}^{(s)}$ \\ \hline
\rule[-10pt]{0pt}{24pt}
$\langle g \rangle$ & 
$4\epsilon$ & $0$
\\
\rule[-10pt]{0pt}{26pt}
$\langle g^2 \rangle$ & 
$\dfrac{8\epsilon}{3}(1 + \epsilon + m)$ & $-\dfrac{4 (4l^2-m^2) \left(l^2+4 q_1^2\right) \left(l^2+4 q_2^2\right)}{3 (2 l+1)} + \{l \mapsto -1-l\}$
\\
\rule[-14pt]{0pt}{38pt}
$\langle g^3 \rangle$ & 
$
\dfrac{8\epsilon}{15}(\epsilon+1)(\epsilon+4)
+ \dfrac{\epsilon m}{15} (36 + 48 \epsilon + 19 m)
$ & $
-\dfrac{2 (4l^2 - m^2) (l^2 + 4q_1^2) (l^2 + 4q_2^2) (16\epsilon + 4l^2 - m^2 + 12)}{5 (2 l+1)}
+ \{l \mapsto -1-l\}
$ \\
\hline
\rule[-12pt]{0pt}{34pt} $F \langle g \rangle$ & 
$\dfrac{4\epsilon}{3}(1 - 2\epsilon + m)$ & $-\dfrac{2 (4l^2 - m^2) \left(l^2+4 q_1^2\right) \left(l^2+4 q_2^2\right)}{3 (2 l+1)} + \{l \mapsto -1-l\}$ \\
\end{tabular}
\end{ruledtabular}
\label{tbl:result_coefs}
\end{table*}

We have thus established the following form of the three-parameter part of the heat kernel valid for any positive $m$:
\begin{equation}
 \Psi^{(3)}
  = \sum_{l=m/2}^\infty \int_{-\infty+im/4-i0}^{\infty+im/4-i0} dq_{1,2}\, \tilde\mu_{\bm{q}}\, \phi_{\bm{q}} e^{-2 \epsilon_{\bm{q}} x/\xi}.
\end{equation}
As was already done earlier for the one-parameter family, we can now shift integration contours for $q_{1,2}$ back to the real axis and collect residues in the poles crossed during this shift. This way we will identify all the relevant eigenstates of the Hamiltonian (\ref{Hamiltonian_radial}) in the three-parameter family. In order to do this, we first change $q_2$ to a new integration variable $q_2 - q_1$ which runs along the real axis. Then we shift the integration contour for $q_1$ picking residues at the poles (\ref{poles21}). Then we restore the $q_2$ variable and in a similar way shift its integration contour to the real axis. In the end we also ``fold'' the double integral domain back to the region $q_1 > q_2 > 0$ and restore the proper integration measure $\mu_q$. This procedure yields
\begin{equation}
 \Psi^{(3)}
  = \sum_{l = m/2}^\infty \Bigl[ \Psi^{(3a)}_l + \Psi^{(3b)}_l + \Psi^{(3c)}_l \Bigr]
\end{equation}
with the following three terms:
\begin{equation}
 \Psi^{(3a)}_l
  = \int_{q_1 > q_2 > 0} \!\!dq_{1,2}\, \mu_{\bm{q}}\, \phi_{\bm{q}}\, e^{-2 \epsilon_{\bm{q}} x/\xi},
\end{equation}
\vskip-4mm
\begin{multline}
 \Psi^{(3b)}_l
  = 8i \sum _{n_1 = 1}^{\lfloor \frac{m}{4} + \frac{1}{2} \rfloor}
        \int_0^\infty \!\!dq_2\, \rho_{\bm{q}} \Bigl[
          \coth\Bigl( \pi q_2 + \frac{i \pi m}{4} \Bigr) \\{} + \coth\Bigl( \pi q_2 - \frac{i \pi m}{4} \Bigr)
        \Bigr] \phi_{\bm{q}}\, e^{-2 \epsilon_{\bm{q}} x/\xi}
        \biggr|_{q_1 = \frac{im}{4}-i n_1 + \frac{i}{2}},
\end{multline}
\vskip-4mm
\begin{equation}
 \Psi^{(3c)}_l
  = 16 \sum _{n_1 = 1}^{\lfloor \frac{m}{4} + \frac{1}{2} \rfloor} \sum_{n_2=n_1}^{\lfloor \frac{m}{4} \rfloor}
    \rho_{\bm{q}}\, \phi_{\bm{q}}\, e^{-2 \epsilon_{\bm{q}} x/\xi} \biggr|_{\substack{q_1 = \frac{im}{4}-i n_1 + \frac{i}{2}, \\ q_2 = \frac{im}{4} - in_2.}}
\end{equation}
We see that aside from the standard continuous set of eigenfunctions with real values of $q_{1,2}$ there are two extra contributions: a subfamily of eigenfunctions with discrete imaginary $q_1$ and continuous positive real $q_2$ and a fully discrete set with both $q_1$ and $q_2$ taking imaginary values. The former subfamily of eigenfunctions is localized only in $\thone$ while the latter functions are localized fully in both noncompact variables $\thone$ and $\thtwo$. The first half-localized eigenfunction appears when $m$ exceeds $2$ while the fully localized eigenstate is possible provided $m \geq 4$.

\section{Results}
\label{sec:results}

Once the expansion of the heat kernel (\ref{Psi}) in eigenfunctions is fully established, we are in a position to evaluate all transport characteristics of the system from Eq.~(\ref{G_through_Z}) with the partition function (\ref{Z_eq_psi}). Equations \eqref{G_through_Z} imply taking partial derivatives at $\theta_i=0$ along the directions corresponding to different $\theta_i$. We cannot calculate the mixed derivatives directly, as we did not obtain the wavefunctions on the whole manifold in an explicit form. However, in Ref.~\onlinecite{classD_first} we argued that it is sufficient to know $\thF$-derivatives only, because the rest can be expressed by applying the Schr\"odinger equation \eqref{Schroedinger_equation} to the small-$\theta$ expansion.
We thus use Eqs.~\eqref{psi1P_fermionic_line} and \eqref{psi3P_fermionic_line} to express conductance moments in the following form:
\begin{multline}
\label{result_general}
 \langle g^s \rangle = \left( \frac{m}{2} \right)^{s} + \int_{-\infty - i/2 + im/4}^{\infty - i/2 + im/4} dq_1\, \tilde\mu_{q_1} \chi_{q_1}^{(s)} e^{-2 \epsilon_{q_1}  L / \xi} \\
  +\sum_{l=m/2}^\infty \int_{-\infty+im/4-i0}^{\infty+im/4-i0} dq_{1,2}\, \tilde\mu_{\bm{q}}\, \chi_{\bm{q}}^{(s)} e^{-2 \epsilon_{\bm{q}}  L/\xi},
\end{multline}
where $\bm{q} = (q_1, q_2, l)$ and the coefficients $\chi_{q_1}^{(s)}$  and $\chi_{\bm{q}}^{(s)}$ are listed in Table~\ref{tbl:result_coefs}. In the same table we provide coefficients for the (pseudo)Fano factor of thermal shot noise power, which is described by a similar integral representation, but without the zero mode contribution [the first term in \eqref{result_general}]. The prefix `pseudo' emphasizes the fact that $F$ is the ratio of independently averaged shot noise power and current (see Ref.\ \onlinecite{classD_first} for more details).

\begin{figure*}
\vskip-2mm
\includegraphics[width=0.95\linewidth]{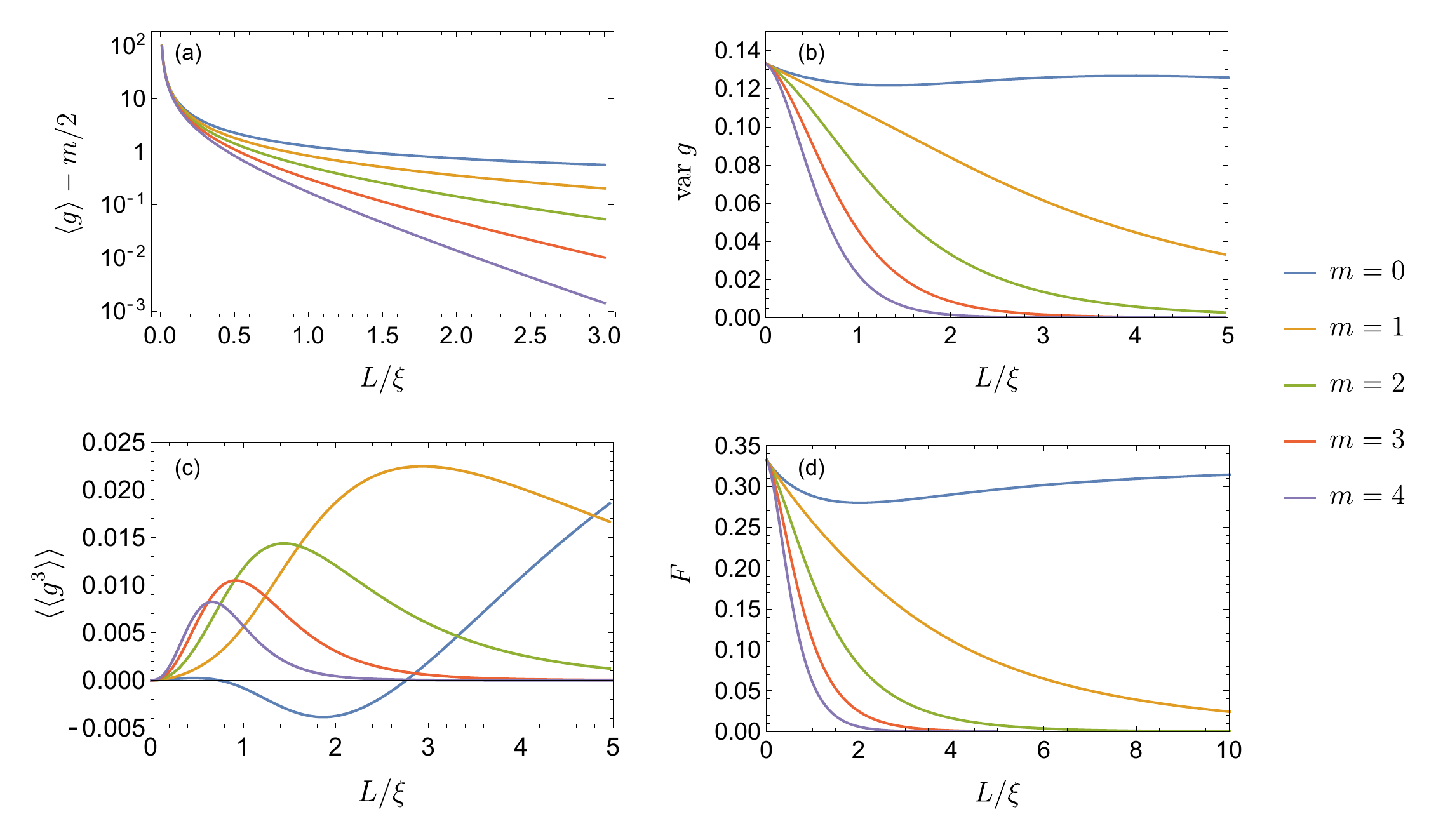}
\vskip-3mm
\caption{Transport characteristics as a function of the system length $L$ in the presence of $m$ topologically protected modes:
	(a)~average conductance of unprotected modes in the logarithmic scale (with the topological contribution $m/2$ subtracted);
	(b)~conductance variance;
	(c) third cumulant of conductance $\langle\langle g^3 \rangle \rangle$;
	(d) thermal Fano factor (ratio of disorder-averaged shot noise to the averaged heat current).}
\label{fig:four_graphs}
\end{figure*}

The first term in \eqref{result_general} stems from the zero mode ($\Psi_0$) part of the heat kernel \eqref{Psi} and describes the contribution of $m$ protected modes. The latter are unidirectional in class D, contributing $m/2$ to the mean conductance averaged over opposite current directions [see the definition \eqref{g_tot}]. At the same time, there is no contribution from the zero mode to the Fano factor because protected modes provide a quantized conductance, which is not subject to the shot noise. Also, according to Table~\ref{tbl:result_coefs} three-parametric eigenfunctions ($\Psi^{(3)}$) do not contribute to the average conductance  (formally since their small-$\theta$ expansion starts from $\theta^4$). This reflects the fact that $\corr{g}$ can be studied with a simpler one-replica sigma-model \cite{KSO2016, Eslam_thesis}.

Equation \eqref{result_general} provides the main result of our paper in the integral form. It contains integrations/summations over Iwasawa momenta that are complicated and probably cannot be taken analytically in the general case. We thus restrict analytic investigation to the asymptotical regions of large/small wire lengths $L$ and complement them with the numerical computation for intermediate $L$. The latter are presented graphically in Figs.~\ref{fig:conductance} and \ref{fig:four_graphs}.

\emph{In the long-wire limit}, $L \gg \xi$, the asymptotic behaviour of conductance moments is determined by eigenfunctions with the lowest eigenvalue $\epsilon_\nu$, since they contribute the slowest exponential to the heat kernel \eqref{Psi}. The case with $m=0$ was discussed in detail in Ref.~\onlinecite{classD_first}. For $ m \leq 4$, the lowest exponential is provided by the one-parametric eigenfunction with nearly zero momentum $q_1 \sim 0$. The corresponding asymptotics can be obtained via the steepest decent method. As long as $m>4$, the lowest eigenvalue is provided by the bound state, which originates from the one-parameter family taken at an imaginary momentum $q_1 = i(m/4 - 1)$ [see Eq.\ \eqref{Psi1}] and has the eigenvalue $\epsilon_{q_1}=(m-2)/2$. Within described approach, we obtain the following asymptotics for conductance moments:
\be
\label{g-gg-long}
	\langle g^s \rangle \sim \left( \frac{m}{2} \right)^s +  \alpha_s^{(m)}
  \begin{cases}
    \displaystyle
     g_L,  & m = 0 ,
\\[4pt]
    \displaystyle
     \frac{\pi^2 g_L \xi e^{-m^2 L/8\xi}}{4 L \sin^2 (\pi m/4)} ,  & 0<m<4,
\\[12pt]
    \displaystyle
    g_L e^{-2 L/\xi}, & m = 4,
\\[6pt]
    \displaystyle
     e^{-(m - 2) L/\xi},  &  m>4 ,
  \end{cases}
\ee
where $g_L=\sqrt{2\xi/\pi L}$ and numerical coefficients $\alpha_s^{(m)}$ are given in Table~\ref{tbl:asymptotics} for different values of $m$.
A similar expression works for the Fano factor but with the first term [$(m/2)^s$] omitted and with the coefficient given in the last line of the same table. For $m>0$, $\corr{g}-m/2\propto e^{-L/L_\text{loc}(m)}$, with the localization length given by Eq.\ \eqref{loc-length}.

\emph{The short-wire limit}, $L \ll \xi$, is most easily accessed via direct perturbative solution of the Schr\"odinger equation for the heat kernel \cite{MMZ}. Corresponding calculation is a straightforward generalization of the calculation in the absence of the WZW term that can be found in Ref.\ \onlinecite{classD_first}. The resulting expansions for $\corr{g}$, $\var g = \langle g^2 \rangle - \gavg^2$, $\corr{\corr{ g^3}} = \langle g^3 \rangle - 3 \langle g^2 \rangle \langle g \rangle + 2 \langle g \rangle^3$, and
$F$ read:
\begin{subequations}
\label{g-com-short}
\begin{gather}
\label{g1-short}
	\gavg = \frac{\xi}{L} + \frac{1}{3}  + \left( \frac{m^2}{12} - \frac{1}{15} \right) \frac{L}{\xi}  + \left( \frac{2}{63} - \frac{m^2}{30} \right) \frac{L^2}{\xi^2} +  \dots,
\displaybreak[0]\\
\label{g2-short}
	\var g = \frac{2}{15} - \frac{8}{315} \frac{L}{\xi} + \left( \frac{136}{4725} - \frac{17 m^2}{630} \right)  \frac{L^2}{\xi^2} + \dots ,
\displaybreak[0]\\
\label{g3-short}
	\corr{\corr{g^3}} = \frac{8}{1485} \frac{L^2}{\xi^2} + \left(\frac{2764 m^2}{155925} - \frac{120704}{6449625} \right) \frac{L^3}{\xi^3} + \dots ,
\displaybreak[0]\\
\label{Fano-short}
	F = \frac{1}{3} - \frac{4}{45} \frac{L}{\xi} + \left( \frac{76}{945} - \frac{m^2}{15} \right) \frac{L^2}{\xi^2} + \dots
\end{gather}
\end{subequations}
In the process of $\var g$ calculation, two leading terms proportional to $1/L^2$ and $1/L$ completely cancel, as expected for universal conductance fluctuations \cite{UCF_Lee_Stone}.

At arbitrary $L$, integrations and summations over Iwasawa momenta in Eq.\ \eqref{result_general} should be performed numerically. The result for the disorder-averaged conductance is presented in Fig.\ \ref{fig:conductance} (it was also obtained in the PhD Thesis of one of the authors \cite{Eslam_thesis}). In this graph one can clearly see the crossover from the Drude regime at $L \ll \xi$ to the peculiar regime at large lengths. The former demonstrates the behaviour typical for a diffusive metal, where the conductance scales as $\gavg \sim \xi / L$ and no signatures of topological protection are visible. At very large lengths, only contribution of the topologically protected modes survives, as expected ($m/2$ for the direction-averaged conductance). Extra contribution of unprotected modes decays exponentially according to the localization scenario. To compare the exponents of this process at different $m$, we plot $\gavg - m/2$ in the logarithmical scale in Fig.~\ref{fig:four_graphs}(a). According to Eq.\ \eqref{loc-length}, the localization length for non-topological modes decreases with increasing the number of topological modes $m$, leading to a quicker decay of the conductance. Finally, at $m = 0$ the wire is in a critical Majorana regime, with the conductance decaying as a power law $\gavg \sim \sqrt{\xi / L}$ \cite{BagretsKamenev,classD_first}.

\begin{table}
\caption{Numerical factor $\alpha_s^{(m)}$ in the large-$L$ asymptotics \eqref{g-gg-long} of conductance moments $\langle g^s \rangle$ and the (pseudo)Fano factor $F$ (in the latter case formulas work for $m>0$; for $m=0$ the factor $F$ does not decay but approaches 1/3).}
\label{tbl:asymptotics}
\begin{ruledtabular}
\begin{tabular}{ccc}
\rule[-5pt]{0pt}{14pt} & $0 \leq m \leq 4$ & $m > 4$ \\
\hline
\rule[-5pt]{0pt}{14pt} $\langle g \rangle$ & $1$ & $m-2$ \\
\rule[-10pt]{0pt}{25pt} $\langle g^2 \rangle$ & $\dfrac{2}{3} \left(1 + m + \dfrac{m^2}{16} \right) $ & $m(m - 4)$ \\
\rule[-10pt]{0pt}{25pt} $\langle g^3 \rangle$ & $\dfrac{8}{15} \left(1 + \dfrac{9m}{3} + \dfrac{43m^2}{64} + \dfrac{3m^3}{32} + \dfrac{m^4}{1024} \right)$ & $\dfrac{3 m^2 (m - 4)}{4}$ \\
 \hline
\rule[-10pt]{0pt}{25pt} $F$ 
& $\dfrac{2}{3m}\left(1 + m - \dfrac{m^2}{8} \right)$ & $\dfrac{2(m-4)}{m} $
\end{tabular}
\end{ruledtabular}
\end{table}

Results for the conductance variance for a broad range of system length $L$ can be found in Fig.\ \ref{fig:four_graphs}(b). In the Ohmic regime at small $L$ the variance takes a universal $m$-independent value of $2/15$, as expected from the theory of universal conductance fluctuations \cite{UCF_Lee_Stone}. At large $L$ the variance drops exponentially for $m \neq 0$ since non-topological modes are suppressed, while the contribution of topological modes does not fluctuate from sample to sample.

In Fig.\ \ref{fig:four_graphs}(c) we present our result for the third cumulant of the conductance $\langle\langle g^3 \rangle\rangle = \langle g^3 \rangle - 3 \langle g^2 \rangle \langle g \rangle + 2 \langle g \rangle^3$. The curve for the $m=0$ case eventually decays $\sim L^{-1/2}$ at very large lengths $L \gtrsim 30$ as was shown in Ref.~\onlinecite{classD_first}. For $m\neq0$ the decay is exponential as expected. Remarkably, for any $m$, $\langle\langle g^3 \rangle\rangle$ behaves as $L^2$ at small lengths because the linear term rather unexpectedly vanishes. A somewhat similar cancellation of the leading contribution was reported in the weak-localization regime for the one-dimensional geometry in Ref.\ \onlinecite{vanRossum1997}.

The (pseudo)Fano factor for various lengths can be found in Fig.~\ref{fig:four_graphs}(d), which is similar to the graph for the variance of conductance. Again, in the short-length limit one observes an $m$-independent finite value $1/3$. This value was predicted for a diffusive metal, described by Dorokhov bimodal distribution of transmission eigenvalues \cite{Dorokhov}. At large lengths, all the curves with $m>0$ decay exponentially for the same reason as for the variance: non-topological modes localize and topological contribution is quantized and thus does not contribute to the shot noise.

\section{Conclusion}
\label{sec:conclusion}

In the present paper we performed an extensive study of superconducting quasi-1D systems of symmetry class D with topologically protected modes. Such systems can arise at the boundary of 2D topological superconductors with broken time-reversal and spin-rotational symmetries. In the text, we suggested a particular way of constructing a system with both topological and regular modes by making an interface between two 2D samples with different values of $\mathbb{Z}$-topological index (Fig.~\ref{fig:edge}). However, our results are not restricted by this particular geometry and can be applied to any 1D system, provided the number of conducting channels is large.

For wires of arbitrary length, we report on exact expressions for the average thermal conductance, its variance, and third cumulant, as well as the shot noise power. The average conductance was evaluated previously in Ref.\ \onlinecite{Eslam_thesis}, where a minimal one-replica version of the sigma-model was used. In our work we resorted to the two-replica nonlinear supersymmetric sigma-model, which allows to access higher moments of transport characteristics.

As was argued in Ref.\ \onlinecite{classD_first}, in principle, one can extract the full counting statistics (FCS) from the two-replica sigma model in class D. Indeed, distribution of transmission probabilities is expressed via the heat kernel in the vicinity of the ``supersymmetric line'' $\thone = \thtwo = - i \thF$. There is no such line in one-replica model due to the degeneracy of the compact sector of the theory. However we did not present explicit results for FCS due to a very complicated structure of the integral representation \eqref{isotropization} in the Iwasawa construction, which prevents us from writing the heat kernel in an analytic form on the supersymmetric line. Instead, we evaluated the heat kernel in the vicinity of the origin $Q = \Lambda$, which suffices to calculate a number of physical quantities mentioned above.

Our findings illustrate in detail the crossover from the ohmic regime at small lengths to the regime of developed localization at large lengths. In particular, we showed that the presence of topologically protected modes enhances localization of unprotected modes. Results for the average conductance $\gavg$ coincide with the results obtained from a simpler (one-replica) sigma-model in Ref.~\onlinecite{Eslam_thesis}. However, as we obtain them from a more complicated two-replica theory, such a coherence is an extra check for our construction.

As we study a superconducting class without a spin-rotational symmetry, experimental verification of our results requires thermal transport measurements. That could be a challenging task, however, recent developments indicate that such measurements are possible at a mesoscopic scale \cite{SET_heat_transport, Khrapai}. To perform disorder-averaging experimentally, one can use a common trick of varying magnetic field or chemical potential in a range that preserves topological state in the system but effectively perturbs the disorder potential.

From a technical perspective, calculation of the sigma-model heat kernel boils down to finding the radial eigenbasis of the transfer-matrix Hamiltonian. In our previous work \cite{classD_first}, we considered the class D sigma model without protected modes, where the Hamiltonian was given by the Laplace-Beltrami operator on the sigma-model supermanifold. Here, topological modes described by an additional WZW term in the sigma-model action introduce a vector potential to the Laplacian. This vector potential has a peculiar form that allows to identify eigenfunctions of the transfer-matrix Hamiltonian in the Iwasawa coordinates in a specially chosen gauge (\ref{T_gauge_choice}). The eigenfunctions are modified by the WZW term and additional bound states arise in the eigenbasis. In our work, we have demonstrated that normalization of the eigenfunctions by their large-$\theta$ asymptotics provides a valid expression in the presence of the WZW term. In the two-replica theory, emerging bound states have a very rich structure. Even in the absence of the WZW term, there are two families of eigenfunctions and a zero mode in the eigenbasis \cite{classD_first}. We showed that both families  give rise to a number of bound states. Moreover, the three-parameter family produces two subclasses of bound states that are localized in either one or two directions in the noncompact sector.

We presented a scheme that allows to identify the bound states algorithmically based on analytic continuation of the heat kernel in the topological index $m$. This approach can be straightforwardly extended to sigma models with arbitrary number of replicas as well as to the sigma models of other symmetry classes. In particular, the scheme can explain the bound states that were found in Ref.\ \onlinecite{MMZ} for the symplectic symmetry class. Our findings extend the theory of Fourier analysis on symmetric superspaces to the case of non-trivial topology and can be useful in a number of physical applications in systems of different symmetry.

\acknowledgments

We are grateful to E.\ J.\ K\"onig for valuable discussions. This work was partially supported by the Russian Science Foundation under Grant No.\ 20-12-00361.

\appendix
\section{Derivation of the sigma model}
\label{appsigma}

In this Appendix we outline a derivation of the sigma-model action to illustrate appearance of the WZW term in Eq.\ \eqref{sigma_model_action}. To simplify some technical details, we will derive a compact replica version of the sigma model.

We start with the 1D Hamiltonian describing $n_L$ left-moving and $n_R$ right-moving channels with a Gaussian disorder potential represented by an antisymmetric matrix in the space of channels:
\begin{gather}
 H
  = -i v \begin{pmatrix} \mathbb{1}_{n_R} & 0 \\ 0 & -\mathbb{1}_{n_L} \end{pmatrix} \frac{\partial}{\partial x} + V(x),
 \quad
 V = - V^T, \label{Horiginal} \\
 \left< V_{ab}(x) V_{cd}(x') \right>
  = \frac{v}{n \tau} \Bigl[ \delta_{ad} \delta_{bc} - \delta_{ac} \delta_{bd}  \Bigr] \delta(x - x'). \label{Gauss_disorder}
\end{gather}
For simplicity, we have assumed the Fermi velocity $v$ to be the same in all channels. Strength of disorder is quantified by the mean free time parameter $\tau$ and also contains the total number of channels $n = n_L + n_R$. Antisymmetry of the Hamiltonian (\ref{Horiginal}) implies the class D.

Fermionic action in terms of Grassmann-valued fields $\phi$ for the above Hamiltonian has the form
\begin{equation}
 S = -i (\phi^\alpha)^\dagger \bigl[ i0 - H \bigr] \phi^\alpha.
\end{equation}
Here we have included an infinitely small imaginary energy $i0$ in order to average a retarded Green function. The replica index $\alpha$ is indicated explicitly and the summation over $\alpha$ is implied. Channel indices are implicit.

Using the antisymmetry of the Hamiltonian we double the variables introducing the PH structure of the fields and define the charge conjugation operation:
\begin{gather}
 S
  = -i \bar\psi^\alpha \bigl[ i0\Lambda - H \bigr] \psi^\alpha, \\
 \psi = \frac{1}{\sqrt{2}} \begin{pmatrix} \phi \\ \phi^* \end{pmatrix},
 \qquad
 \bar\psi = \frac{1}{\sqrt{2}} \Bigl( \phi^\dagger,\; \phi^T \Bigr) = \psi^T C,
 \\
 C = \begin{pmatrix} 0 & 1 \\ 1 & 0  \end{pmatrix},
 \qquad
 \Lambda = \begin{pmatrix} 1 & 0 \\ 0 & -1 \end{pmatrix}.
\end{gather}
Charge conjugation obeys the following antisymmetry property for any two vectors $\psi_{1,2}$:
\begin{equation}
 \bar\psi_1 \psi_2 = \psi_1^T C \psi_2 = -\psi_2^T C \psi_1 = -\bar\psi_2 \psi_1.
 \label{Csym}
\end{equation}

Averaging the action with respect to the Gaussian disorder (\ref{Gauss_disorder}), we encounter the following four-fermionic term [here both replica ($\alpha$, $\beta$) and channel ($a$, $b$) indices are explicitly shown]:
\begin{equation}
 \left< \bar\psi_a^\alpha V_{ab} \psi_b^\alpha \bar\psi_c^\beta V_{cd} \psi_d^\beta \right>
  = -\frac{2v}{n \tau} \operatorname{tr} \Bigl(\psi_a^\beta \bar\psi_a^\alpha \psi_b^\alpha \bar\psi_b^\beta \Bigr).
\end{equation}
In deriving the last expression, we have used the identity (\ref{Csym}) and represented the result as a trace in the PH space. Replica indices are arranged such that we can extend the trace operation to the replica space. Then it will become a trace of the square of the matrix $\psi_a \bar\psi_a$ which is already summed over the repeated channel index.

We can decouple the four-fermion term with the help of Hubbard-Stratonovich transformation by introducing an auxiliary matrix field $Q$ that acts in the replica and PH spaces only
\begin{multline}
 S
  = \frac{n}{16 v \tau} \operatorname{Tr} Q^2 \\ - i \bar\psi^\alpha \left[
      i0\Lambda + i v \begin{pmatrix} \mathbb{1}_{n_R} & 0 \\ 0 & -\mathbb{1}_{n_L} \end{pmatrix} \frac{\partial}{\partial x} + \frac{i Q}{2\tau}
    \right] \psi^\alpha.
 \label{Spsi}
\end{multline}
Now the action is again quadratic in $\psi$ and the operator in square brackets is almost trivial in the space of channels. The only remaining dependence is in the sign of the kinetic term. Let us also point out that the charge conjugation relation between $\psi$ and $\bar\psi$ effectively restricts the matrix $Q$ as
\begin{equation}
 Q = - C^T Q^T C = - \bar Q.
 \label{QbarQ}
\end{equation}
Gaussian integration over $\psi$ with the action (\ref{Spsi}) yields a square root of the determinant of the operator in brackets. The action is then
\begin{multline}
 S
  = \frac{n}{16 v \tau} \operatorname{Tr} Q^2
    - \frac{n_R}{2} \operatorname{Tr} \ln \left[ i 0 \Lambda + i v \frac{\partial}{\partial x} + \frac{i Q}{2\tau} \right] \\
    - \frac{n_L}{2} \operatorname{Tr} \ln \left[ i 0 \Lambda - i v \frac{\partial}{\partial x} + \frac{i Q}{2\tau} \right].
 \label{Strln}
\end{multline}
At this stage we have completely eliminated the channel space. All remaining traces operate only in the PH and replica spaces.

Saddle point analysis of the action (\ref{Strln}) yields $Q = \pm 1$. The infinitely small term $i0\Lambda$ in the argument of the logarithm indicates a proper arrangement of signs on the diagonal of $Q$. We thus pick the saddle point solution $Q = \Lambda$ and drop the term $i0\Lambda$ from the action altogether. Other possible saddle points can be generated by rotations of $\Lambda$ hence the sigma-model manifold is parametrized as
\begin{equation}
 Q
  = T^{-1} \Lambda T
 \label{QTLambdaT}
\end{equation}
such that the constraint (\ref{QbarQ}) is preserved [cf.\ Eq.\ (\ref{Q_via_Lambda_and_T})].

The matrix $T$ has the size $2n$, with $n$ being the number of replicas, and obeys $\bar T T = 1$. The space of $T$ is the group $\mathrm{SO}(2n)$. (We disregard matrices $T$ with a negative determinant and consider only the connected part of the sigma-model manifold as explained in the main text.) The matrix $Q$ does not change if we replace $T \mapsto KT$ with any matrix $K$ that commutes with $\Lambda$. Such matrices form a subgroup of $\mathrm{SO}(2n)$ that has the structure of $\mathrm{U}(n)$. We thus conclude that matrix $Q$ belongs to the coset space $\mathrm{SO}(2n) / \mathrm{U}(n)$ as it should be for class D.

To derive the sigma-model action on the saddle manifold (\ref{QTLambdaT}), we expand the action (\ref{Strln}) in gradients assuming $Q$ is slowly varying in space. Usually, this gradient expansion is done in terms of the matrix $T$. We will take an alternative approach and perform the gradient expansion directly in terms of $Q$ to maintain an apparent gauge symmetry of the theory. In order to do so, we first extend the definition of the matrix $Q$ in the auxiliary direction $s$ such that
\begin{equation} \label{Qsx}
 Q(s,x)
  = \begin{cases}
     Q(x), & s = 1, \\
     \Lambda, & s = 0.
    \end{cases}
\end{equation}
The extended matrix $Q$ is a smooth function of its both arguments $s$ and $x$ and interpolates between the physical value $Q(x)$ for $s = 1$ and a chosen point $\Lambda$ for $s = 0$. The physical field $Q(x)$ represents a closed curve on the sigma-model manifold. Hence the extension (\ref{Qsx}) is always possible since the manifold $\mathrm{SO}(2n) / \mathrm{U}(n)$ is simply connected.

Let us consider the first logarithmic term in the action (\ref{Strln}). Using the extended version of the matrix $Q$, we write this term in the integral form
\begin{equation} \label{Trln_int_s}
 \operatorname{Tr} \ln \left( \frac{i Q}{2\tau} - v p \right)
  = \frac{i}{2\tau} \int_0^1 ds \operatorname{Tr} \frac{\partial Q}{\partial s} \left( \frac{i Q}{2\tau} - v p \right)^{-1}.
\end{equation}
Here $p = -i\partial/\partial x$ is the standard momentum operator. Now, instead of the logarithm, we should expand an inverse operator in gradients of $Q$.  This can be done with the help of the following identity:
\begin{multline} \label{GQ}
 \left( \frac{i Q}{2\tau} - v p \right)^{-1} \\
  = -\left( \frac{i Q}{2\tau} + v p \right) \left( v^2 p^2 + \frac{1}{4\tau^2} - \frac{v}{2 \tau} \frac{\partial Q}{\partial x} \right)^{-1} \\
  \approx -\left( \frac{i Q}{v} + 2\tau p \right) \left( g + g \frac{\partial Q}{\partial x} g + g \frac{\partial Q}{\partial x} g \frac{\partial Q}{\partial x} g \right).
\end{multline}
Here we have introduced the notation
\begin{equation}
  g = \left(2 l p^2 + \frac{1}{2l} \right)^{-1}.
\end{equation}
The operator $g$ is diagonal in momentum representation and decays at the mean free path scale $l = v\tau$ in real space.

The inverse operator in the integrand of Eq.\ (\ref{Trln_int_s}) is taken at coincident points. In order to calculate it, we need to commute all $\partial Q/\partial x$ factors in Eq.\ (\ref{GQ}) to the right and then integrate over $p$. Commutation is performed using the identity
\begin{equation} \label{g_comm}
 A g - g A
  = -2 i l g \left( p \frac{\partial A}{\partial x} + \frac{\partial A}{\partial x} p \right) g
\end{equation}
that holds for any $x$-dependent function $A$. We see that commutation with $g$ generates derivatives in $x$. Hence in the very last term of Eq.\ (\ref{GQ}), we can simply interchange the order of factors and neglect the commutator as long as we keep only terms with up to two derivatives in $x$. For the second to last term, we will apply the identity (\ref{g_comm}) once and then rearrange the factors neglecting higher derivatives. This yields
\begin{multline} \label{GQxx}
 \left( \frac{i Q}{2\tau} - v p \right)^{-1}_{x,x}
  = -\int \frac{dp}{2\pi v} \bigl( i Q + 2 l p \bigr) \\
    \times \left[
      g + g^2 \frac{\partial Q}{\partial x} - 4 i l p g^3\, \frac{\partial^2 Q}{\partial x^2} + g^3 \left( \frac{\partial Q}{\partial x} \right)^2
    \right] \\
  = -\frac{i}{2v} \left[
      Q + l Q \frac{\partial Q}{\partial x} - l^2\, \frac{\partial^2 Q}{\partial x^2} + \frac{3 l^2}{2} Q \left(\frac{\partial Q}{\partial x}\right)^2
    \right].
\end{multline}
We substitute this result into Eq.\ (\ref{Trln_int_s}) and observe that the first and the last terms drop out under the trace since $Q$ anticommutes with any derivative of $Q$. The second to last term in Eq.\ (\ref{GQxx}) can be integrated by parts in $x$ which then allows to integrate it over $s$. Gradient expansion then takes the form
\begin{multline} \label{Trln_expanded}
 \operatorname{Tr} \ln \left( \frac{i Q}{2\tau} - v p \right) \\
  = \frac{l}{8} \operatorname{Tr} \left( \frac{\partial Q}{\partial x} \right)_{s = 1}^2
    + \frac{1}{4}\int_0^1 ds \operatorname{Tr} \left( \frac{\partial Q}{\partial s} Q \frac{\partial Q}{\partial x} \right).
\end{multline}

The first term in this result represents the standard kinetic term of the sigma-model action. It explicitly depends only on the physical value of $Q$ at $s = 1$. The last term in Eq.\ (\ref{Trln_expanded}) is more subtle. It has the standard Wess-Zumino-Witten form defined in terms of the extended field $Q(s,x)$. On the other hand, it also has the form of the Pruisken topological term for the sigma model of a quantum-Hall system in the $(s,x)$ 2D plane. The physical 1D space with $s = 1$ represents a boundary to this fictitious 2D quantum-Hall system. In order to rewrite the WZW term as a function of only the physical field, we resort to the parametrization (\ref{QTLambdaT}). This allows us to write
\begin{multline} \label{WZWQT}
 \int_0^1 ds \operatorname{Tr} \left( \frac{\partial Q}{\partial s} Q \frac{\partial Q}{\partial x} \right) \\
  = 2 \int_0^1 ds \operatorname{Tr} \left[
      \frac{\partial}{\partial s} \left( T^{-1} \Lambda \frac{\partial T}{\partial x} \right)
      -\frac{\partial}{\partial x} \left( T^{-1} \Lambda \frac{\partial T}{\partial s} \right)
    \right] \\
  = 2 \operatorname{Tr} \left( T^{-1} \Lambda \frac{\partial T}{\partial x} \right)_{s = 1}.
\end{multline}
The gradient expansion now takes its final form
\begin{equation}
 \operatorname{Tr} \ln \left( \frac{i Q}{2\tau} - v p \right)
  = \frac{l}{8} \operatorname{Tr} \left( \frac{\partial Q}{\partial x} \right)^2
    + \frac{1}{2} \operatorname{Tr} \left( T^{-1} \Lambda \frac{\partial T}{\partial x} \right).
\end{equation}
Applying the same calculation strategy to the last term of Eq.\ (\ref{Strln}) we obtain the complete sigma-model action
\begin{equation} \label{Ssigma}
 S
  = \int dx \operatorname{tr} \left[
      \frac{n l}{16} \left(\frac{\partial Q}{\partial x}\right)^2
      + \frac{m}{4}\, T^{-1} \Lambda \frac{\partial T}{\partial x}
    \right].
\end{equation}
It exactly reproduces the compact part of the supersymmetric action (\ref{sigma_model_action}) from the main text.

Let us comment on the gauge invariance of this result. The WZW term in the right-hand side of Eq.\ (\ref{Ssigma}) is written in terms of $T$ rather than $Q$. It may seem that a transformation $T \mapsto KT$ with $[\Lambda, K] = 0$, that does not alter $Q$, may change the action of the model. This is actually not the case provided the following two conditions are in place: (i) the coefficient in front of the WZW term is an integer multiple of $1/4$, (ii) the field configuration $T(x)$ is topologically trivial.

The first condition is required to ensure that the topological term in Eq.\ (\ref{Trln_expanded}) is well-defined. For the same physical field $Q(x)$ there are topologically distinct extensions (\ref{Qsx}) to the second direction $s$. While the topological term is invariant with respect to small variations of $Q$, it can take different values for topologically different configurations of $Q(s,x)$. When the prefactor of the topological term is properly quantized, the action for topologically distinct configurations will differ by an integer multiple of $2\pi i$ and hence the partition function of the model will not depend on the particular choice of extension. For our problem, this quantization condition is fulfilled automatically since the difference $n_R - n_L$ is always integer.

The second condition ensures that the transformation (\ref{WZWQT}) can be safely performed. Indeed, in Eq.\ (\ref{WZWQT}), we apply the parametrization (\ref{QTLambdaT}) to the whole extended field $Q(s,x)$. The sigma-model manifold [coset space $\mathrm{SO}(2n) / \mathrm{U}(n)$] is simply connected, which allows to build the extension (\ref{Qsx}) for any physical field $Q(x)$. However, this is not the case for the group $\mathrm{SO}(2n)$ to which the matrix $T$ belongs. This means that in order to have a smooth extended field $T(s,x)$ we should require that the parametrization $T(x)$ of the physical field $Q(x)$ is chosen such that it can be continuously deformed to a single point, that is $T(x)$ belongs to the trivial homotopy class of the group $\mathrm{SO}(2n)$. Then a continuous transformation of $Q(x)$ at $s = 1$ to $Q = \Lambda$ at $s = 0$ can be parametrized by the corresponding smooth transformation of $T(x)$ at $s = 1$ to $T = 1$ at $s = 0$. With this condition fulfilled, the gauge symmetry of the action becomes obvious. Indeed, the WZW term is directly identified with the Pruisken topological term by Eq.\ (\ref{WZWQT}) while the latter is written explicitly in terms of $Q$ and hence is gauge invariant.

Let us also analyze how the action changes under a gauge transformation. Consider a transformation
\begin{equation} \label{gauge}
 T \mapsto U T
\end{equation}
with a matrix $U$ that obeys $\bar U U = 1$ and commutes with $\Lambda$. The latter condition implies $U$ is block-diagonal, while the former condition establishes a relation between the blocks:
\begin{equation}
 U = \begin{pmatrix} u & 0 \\ 0 & u^* \end{pmatrix}
\end{equation}
with $u$ being an arbitrary unitary matrix of size $n$. The transformation (\ref{gauge}) leaves $Q$ invariant but changes the action (\ref{Ssigma}) as
\begin{equation}
\label{gauge_action}
 S
  \mapsto S + \frac{m}{4} \int_0^L dx \operatorname{tr} \left( U^{-1} \Lambda \frac{\partial U}{\partial x} \right)
  = S + \frac{m}{2} \operatorname{tr} \ln u \Bigr|_0^L.
\end{equation}
We assume that the trajectory $T(x)$ is closed and hence $U(x)$ also must be closed. Determinant of $u$ can change its phase by any integer multiple of $2\pi$ along such a closed trajectory. If $m$ is even, the action is also shifted by an integer multiple of $2\pi i$ under the gauge transformation and hence the theory is fully gauge invariant.

For odd values of $m$, gauge invariance of the theory is enforced provided $\ln\det u$ changes by a multiple of $4\pi i$. Earlier, we have established an additional condition on the allowed trajectories of $T(x)$: the closed path $T(x)$ should be possible to continuously shrink to a single point. The fundamental group of the manifold of $T$ is $\pi_1(\mathrm{SO}(n)) = \mathbb{Z}_2$. So there are only two topologically distinct classes of closed trajectories: those that can be shrunk to a point and those that cannot. It is straightforward to check that a gauge transformation (\ref{gauge}) with $\ln\det u$ changing by an odd multiple of $2\pi i$ maps these two classes on to each other. Hence we conclude that our requirement on the trajectory of $T$ to be topologically trivial rules out such odd gauge transformations and indeed enforces the gauge invariance of the theory.

\section{Evaluation of the gauge factor}
\label{app:gauge}

In this Appendix, we explain how to resolve Eq.\ (\ref{gauge_change}) and express the gauged plane wave \eqref{gauged_wavefunction} in terms of Cartan variables $\theta_i$ and $U$. This result will be used to evaluate the isotropization formula \eqref{isotropization}.

In the main text of the paper, we use a representation of the sigma model that renders Cartan subalgebra matrices $\check\theta$ explicitly diagonal:
\begin{equation} \label{diag_theta}
 \check\theta
  = \operatorname{diag}\{\thone, \thtwo, i\thF, i\thF, - i\thF, -i\thF, -\thtwo, -\thone \}.
\end{equation}
At the same time, $\Lambda$ is a unit antidiagonal matrix and hence anticommutes with $\hat\theta$. Both $\Lambda$ and $\hat\theta$ change sign under charge conjugation with the matrix
\begin{equation} \label{C}
 C
  = \begin{pmatrix}
      0 & 0 & 0 & 0 & 0 & 0 & 0 & 1 \\
      0 & 0 & 0 & 0 & 0 & 0 & 1 & 0 \\
      0 & 0 & 0 & 0 & 1 & 0 & 0 & 0 \\
      0 & 0 & 0 & 0 & 0 & -1 & 0 & 0 \\
      0 & 0 & 1 & 0 & 0 & 0 & 0 & 0 \\
      0 & 0 & 0 & -1 & 0 & 0 & 0 & 0 \\
      0 & -1 & 0 & 0 & 0 & 0 & 0 & 0 \\
      -1 & 0 & 0 & 0 & 0 & 0 & 0 & 0
    \end{pmatrix},
\end{equation}

We start with the identity (\ref{gauge_change}) between the Cartan and Iwasawa decompositions of the matrix $T$:
\be
	\label{equating_T}
	T = U^{-1} e^{\check{\theta}/2} U = V e^{\check{a} / 2} N.
\ee
Let us remind that $U$ and $V$ belong to the $K$ group and commute with $\Lambda$, $\check{\theta}$ and $\check{a}$ are Cartan and Iwasawa radial angles, respectively, represented by diagonal matrices of the form (\ref{diag_theta}), and $N = e^n$ appears in the Iwasawa decomposition \eqref{T_Iwasawa} and is an upper-triangular matrix with units on the main diagonal. We can represent all these matrices explicitly with the following block structure:
\begin{subequations}
\label{app_definitions}
\begin{align}
\check{\theta} &= \begin{pmatrix} \tilde{\theta} & 0 \\ 0 & -R \tilde{\theta} R \end{pmatrix},
&
\check{a} &= \begin{pmatrix} \tilde{a} & 0 \\ 0 & -R \tilde{a} R \end{pmatrix} ,
\\
\Lambda &= \begin{pmatrix} 0 & R \\ R & 0 \end{pmatrix},
&
e^{\check{a} / 2} N &= \begin{pmatrix} X & Y \\ 0 & Z  \end{pmatrix}.
\end{align}
\end{subequations}
Here $R$ is a $4 \times 4$ antidiagonal unit matrix while $\tilde\theta$ and $\tilde a$ are diagonal matrices of the radial Cartan and Iwasawa angles, respectively:
\begin{gather}
 \tilde{\theta} = \diag\,\{\thone, \thtwo,i\thF, i\thF \}, \\
 \tilde{a} = \diag\,\{ \aone, \atwo, i\aF, i\aF \}.
\end{gather}
The blocks $X$, $Y$ and $Z$ contain Iwasawa coordinates and constitute an upper triangular matrix $e^{\check{a}/2} N$. So, the blocks $X$ and $Z$ are themselves upper triangular and contain exponentials of Iwasawa angles $e^{a_i / 2}$ on the main diagonal.

To resolve the identity (\ref{equating_T}) with respect to $V$, we change matrix basis such that $\Lambda$ becomes a diagonal matrix. This can be achieved by an orthogonal rotation with the following matrix $L$:
\be
	L = \frac{1}{\sqrt{2}} \begin{pmatrix} \mathbb{1} & \mathbb{1} \\ R & -R  \end{pmatrix} ,
	\qquad L^T \Lambda L = \begin{pmatrix} \mathbb{1} & 0 \\ 0 & -\mathbb{1} \end{pmatrix} .
\ee
In the new basis, the charge conjugation matrix $C$ [see Eq.\ (\ref{C})] takes a block-offdiagonal form
\begin{equation} \label{Cc}
 L^T C L
  = \begin{pmatrix}
     0 & -c \\
     c & 0
    \end{pmatrix},
 \qquad
 c
  = \begin{pmatrix}
      1 & 0 & 0 & 0 \\
      0 & 1 & 0 & 0 \\
      0 & 0 & 0 & 1 \\
      0 & 0 & -1 & 0
    \end{pmatrix}.
\end{equation}
Since matrices $U$ and $V$ commute with $\Lambda$, they take a block-diagonal form in the new basis:
\begin{equation} \label{LTUL}
 L^T U L = \begin{pmatrix} u & 0 \\ 0 & \bar u^{-1} \end{pmatrix},
 \qquad L^T V L = \begin{pmatrix} v & 0 \\ 0 & \bar v^{-1} \end{pmatrix}.
\end{equation}
Both matrices obey the condition $\bar U U = \bar V V = 1$. This establishes a relation between diagonal blocks of $U$ and $V$ in the new basis as indicted in Eq.\ (\ref{LTUL}). Charge conjugation for individual blocks is defined as $\bar u = c^T u^T c$ with the matrix $c$ from Eq.\ (\ref{Cc}). Representation (\ref{LTUL}) shows explicitly that the matrices $U$ and $V$ can be fully parametrized by their upper-left blocks and hence belong to the supergroup $\mathrm{U}(2|2)$. Finally, the diagonal Cartan matrix $e^{\check\theta}$ acquires the following form in the rotated basis:
\begin{equation}
 L^T e^{\check\theta} L
  = \begin{pmatrix} \cosh\tilde\theta & \sinh\tilde\theta \\ \sinh\tilde\theta & \cosh\tilde\theta \end{pmatrix}.
\end{equation}

The gauge factor in Eq.\ (\ref{gauge_change}) takes especially simple form after $L$ rotation,
\be \label{gauge_v}
  e^{(m/4) \str (\Lambda \ln V)} = \sdet^{m/4} v\, \sdet^{-m/4} \bar v^{-1} = \sdet^{m/2} v.
\ee
To find $\sdet v$, we will rewrite identity (\ref{equating_T}) in the rotated basis and take a superdeterminant of its upper-left block. This yields
\begin{multline}
 \sdet\cosh(\tilde\theta/2)
  = \sdet v\, \sdet(X + Y R + R Z R) \\
  = \frac{\sdet v}{\sdet e^{\tilde{a}/2}} \sdet(1 + R Z^{-1} R X + R Z^{-1} R Y R).  \label{sdet_cosh_theta}
\end{multline}
In the last expression we have singled out the factor $\sdet(RZR) = \sdet e^{-\tilde a/2}$.

We have thus related $\sdet v$ to the Iwasawa coordinates contained in $X$, $Y$, and $Z$. To express these parameters through Cartan coordinates, consider the matrix $\Lambda Q$ in the original basis written in two alternative coordinate systems,
\begin{multline}
 \Lambda Q
  = \Lambda N^{-1} e^{-\check a/2} \Lambda e^{\check a/2} N \\
  = \begin{pmatrix} R Z^{-1} R X & R Z^{-1} R Y \\ -R X^{-1} Y Z^{-1} R X & R X^{-1} R Z - R X^{-1} Y Z^{-1} R Y  \end{pmatrix} \\
  = U^{-1} e^{\check\theta} U
  = L \begin{pmatrix} u^{-1} \cosh\tilde\theta u & u^{-1} \sinh\tilde\theta \bar u^{-1} \\ \bar u \sinh\tilde\theta u & \bar u \cosh\tilde\theta \bar u^{-1} \end{pmatrix} L^T.
\end{multline}
We observe that the blocks in the upper row of this matrix are exactly those appearing in Eq.\ (\ref{sdet_cosh_theta}). Taking these blocks from the matrix $\Lambda Q$ in Cartan coordinates, we express $\sdet v$ as
\begin{equation}
 \sdet v
  = \frac{\sdet e^{\tilde{a}/2}}{\sdet\bigl[\cosh(\tilde\theta/2) + u \bar u \sinh(\tilde\theta/2)\bigr]}.
\end{equation}
This fully determines the phase factor in Eq.\ (\ref{isotropization}) in terms of Cartan angles according to Eq.\ (\ref{gauge_v}).

With this result, we can rewrite the integral representation (\ref{isotropization}) of a general eigenfunction of the transfer-matrix Hamiltonian in the following form:
\be \label{isotropization2}
  \phi_{\bm{q}} (\check\theta)
  = \left<
      \frac{e^{\left(i q_1 + \frac{m}{4} \right) \aone + \left( i q_2 + \frac{m}{4} + \frac{1}{2} \right) \atwo + i \left( l - \frac{m}{2} \right) \aF}}
           {\sdet^{m/2} \bigl[\cosh(\tilde\theta/2) + u \bar u \sinh(\tilde\theta/2)\bigr]}
    \right>_u.
\ee
It remains to express the Iwasawa angles $\check a$ via Cartan coordinates in order to perform isotropization. This can be done by considering superdeterminants of square submatrices of $\Lambda Q$ in the upper-left corner as was discussed in detail in Ref.\ \onlinecite{classD_first}.

\bibliography{class_D_variance}

\end{document}